\documentclass[iop,openany]{emulateapj_14}  
\pdfoutput=1  							 
\usepackage{amsmath}
\usepackage{amssymb}
\usepackage{amstext}  
\usepackage{amsfonts}              
\usepackage[parfill]{parskip}  
\usepackage{graphics}
\usepackage{graphicx}
\usepackage{natbib}
\usepackage{floatrow}
\usepackage{booktabs}

\usepackage{booktabs} 	
\usepackage{float}  		

\usepackage{mathtools}  				
\DeclarePairedDelimiter\abs{\lvert}{\rvert}	 	 
\DeclarePairedDelimiter\norm{\lVert}{\rVert}  	 
\makeatletter
\let\oldabs\abs
\def\abs{\@ifstar{\oldabs}{\oldabs*}}
\let\oldnorm\norm
\def\norm{\@ifstar{\oldnorm}{\oldnorm*}}
\makeatother

\usepackage{relsize}	

\usepackage[usenames,dvipsnames]{color}

\usepackage{epstopdf} 						
\usepackage{epsfig} 						

\usepackage[colorlinks=true,	linkcolor=Red,	urlcolor=blue, citecolor=RoyalBlue, filecolor=magenta anchorcolor=blue runcolor=blue]{hyperref}		
\usepackage{hypernat} 						
\usepackage{etex} 							
\usepackage{bookmark} 						
\usepackage[hang, flushmargin]{footmisc}

\def\farcs{\hbox{$.\!\!^{\prime\prime}$}}
\def\fs{\hbox{$.\!\!^{\rm s}$}}

\newcommand*\slfrac[2]{\left.#1\middle/#2\right.}

\renewcommand{\thempfootnote}{\arabic{mpfootnote}}
\slugcomment{Accepted by ApJ,  2014 September 26}
\shortauthors{\small{Zanardo et al.}} 
\shorttitle{\small{Spectral and morphological analysis of SNR 1987A with ALMA \& ATCA}}

\begin{document}

\title{Spectral and morphological analysis of the remnant of Supernova 1987A \\ with ALMA  \& ATCA}

\author{
Giovanna Zanardo\altaffilmark{1}, 
Lister Staveley-Smith\altaffilmark{1,2}, 
Remy Indebetouw\altaffilmark{3,4}, 
Roger A. Chevalier\altaffilmark{3},
Mikako Matsuura\altaffilmark{5},
Bryan M. Gaensler\altaffilmark{2,6}, 
Michael J. Barlow\altaffilmark{5},
Claes Fransson\altaffilmark{7}, 
Richard N. Manchester\altaffilmark{8},
Maarten Baes\altaffilmark{9},\\
Julia R. Kamenetzky\altaffilmark{10},
Ma\v{s}a Laki\'cevi\'c\altaffilmark{11},
Peter Lundqvist\altaffilmark{7},
Jon M. Marcaide\altaffilmark{12,13},\\
Ivan  Mart\'i-Vidal\altaffilmark{14},
Margaret Meixner\altaffilmark{15,16},
C.-Y. Ng\altaffilmark{17},
Sangwook Park\altaffilmark{18},\\
George Sonneborn\altaffilmark{19},
Jason Spyromilio\altaffilmark{20}, 
Jacco Th. van Loon\altaffilmark{11}
}

\affil{
 \altaffilmark{1}International Centre for Radio Astronomy Research (ICRAR), M468, University of Western Australia, \\Crawley, WA 6009, Australia; \href{mailto:giovanna.zanardo@gmail.com}{giovanna.zanardo@gmail.com}\\
 \altaffilmark{2}Australian Research Council, Centre of Excellence for All-sky Astrophysics (CAASTRO)\\
 \altaffilmark{3}Department of Astronomy, University of Virginia, PO Box 400325, Charlottesville, VA 22904, USA\\
 \altaffilmark{4}National Radio Astronomy Observatory (NRAO), 520 Edgemont Rd, Charlottesville, VA 22903, USA\\
 \altaffilmark{5}Department of Physics and Astronomy, University College London, Gower St., London WC1E 6BT, UK\\
 \altaffilmark{6}Sydney Institute for Astronomy, School of Physics, University of Sydney, NSW 2006, Australia\\ 
 \altaffilmark{7}Department of Astronomy, Oskar Klein Center, Stockholm University, AlbaNova, SE-106 91 Stockholm, Sweden\\
 \altaffilmark{8}CSIRO Astronomy and Space Science, Australia Telescope National Facility, PO Box 76, Epping, NSW 1710, Australia\\
 \altaffilmark{9}Sterrenkundig Observatorium, Universiteit Gent, Krijgslaan 281 S9, B-9000 Gent, Belgium\\
 \altaffilmark{10}Steward Observatory, University of Arizona, 933 North Cherry Avenue, Tucson, AZ 85721-0065, USA\\
 \altaffilmark{11}Institute for the Environment, Physical Sciences and Applied Mathematics, Lennard-Jones Laboratories, \\Keele University, Staffordshire ST5 5BG, UK\\
 \altaffilmark{12}Departamento de Astronom\'ia, Universidad de Valencia, C/Dr. Moliner 50, 46100, Burjassot, Spain\\
 \altaffilmark{13}Donostia International Physics Center, Paseo de Manuel de Lardizabal 4, E-20018 Donostia-San Sebastian, Spain\\
 \altaffilmark{14}Department of Earth and Space Sciences, Chalmers University of Technology, Onsala Space Observatory, SE-439 92 Onsala, Sweden\\					%
 \altaffilmark{15}Space Telescope Science Institute, 3700 San Martin Drive, Baltimore, MD 21218, USA\\
 \altaffilmark{16}Department of Physics and Astronomy, Johns Hopkins University, 366 Bloomberg Center, 3400 N. Charles Street, \\Baltimore, MD 21218, USA\\
 \altaffilmark{17}Department of Physics, University of Hong Kong, Pokfulam Road, Hong Kong, China\\
 \altaffilmark{18}Department of Physics, University of Texas at Arlington, 108 Science Hall, Box 19059, Arlington, TX 76019, USA\\
 \altaffilmark{19}NASA Goddard Space Flight Center, 8800 Greenbelt Rd., Greenbelt, MD 20771, USA\\
 \altaffilmark{20}European Southern Observatory (ESO), Karl-Schwarzschild-Str. 2, 85748 Garching b. M\"{u}nchen, Germany
}

\vspace{0.0mm}
\begin{abstract}
We present a comprehensive spectral and morphological analysis of the remnant of Supernova (SN) 1987A with the Australia Telescope Compact Array (ATCA) and the  Atacama Large Millimeter/submillimeter Array (ALMA). 
The non-thermal and thermal components of the radio emission are investigated in images from 94 to 672 GHz ($\lambda$ 3.2 mm to 450 $\mu$m),  with the assistance of a high-resolution 44 GHz synchrotron template from the ATCA, and a dust template from ALMA observations at 672 GHz.
An analysis of the emission distribution over the equatorial ring in images from 44 to 345 GHz highlights a gradual decrease of the east-to-west asymmetry ratio with frequency. 
We attribute this to the shorter synchrotron lifetime at high frequencies.
Across the transition from radio to far infrared, both the synchrotron/dust-subtracted images  and the spectral energy distribution (SED) suggest additional emission beside the main synchrotron component ($S_{\nu}\propto\nu^{-0.73}$) and the thermal component originating from dust grains at $T\sim 22$ K. 
This excess could be due to free-free flux or emission from grains of colder dust. However, a second flat-spectrum synchrotron component
appears to better fit the SED, implying that the emission could be attributed to a pulsar wind nebula (PWN).
The residual emission is mainly localised west of the SN site, as the spectral analysis yields  $-0.4\lesssim\alpha\lesssim-0.1$ across the western regions, with $\alpha\sim0$ around the central region.
If there is a PWN in the remnant interior, these data suggest that the pulsar may be offset westward from the SN position.
\end{abstract}


\section{Introduction}
\label{Intro}

The evolution of the remnant of Supernova (SN) 1987A in the Large Magellanic Cloud has been closely monitored since the collapse of its progenitor star, Sanduleak (Sk) $-69^{\circ} 202$, on 1987 February 23. 
Models of Sk $-69^{\circ} 202$ indicated that it had an initial mass of $\sim$20 $M_{\sun}$ \citep{hil87}. 
The mass range of the progenitor is consistent with  the formation of a neutron star  \citep{thi85}, and thus with the neutrino events reported by the KamiokaNDE \citep{hir87} and IMB (\citealp{bio87,hai88}) detectors. 
Models by \citet{cro00} suggest a transition from red supergiant (RSG) into blue supergiant (BSG) to explain the hourglass nebula structure, which envelopes the SN with three nearly-stationary rings (\citealp{che95,blo93, mar95,mor07}). 
The two outer rings, imaged with the {\it Hubble Space Telescope} ({\it HST}; \citealp{jak91,pla95}), are located on either side of the central ring in the equatorial plane  ({\it equatorial ring}, ER), and
 likely formed at the same time as the ER (\citealp{cro00,tzi11}). 
The synchrotron emission, generated by the shock propagating into the clumpy circumstellar medium (CSM) close to the equatorial plane, was detected in the mid-90s (\citealp{tur90,sta92}), and has become brighter over time (\citealp{man02,zan10}).
Radio observations have stretched from flux monitoring at 843 MHz with the Molonglo Observatory Synthesis Telescope (\citealp{sta93,bal01}) to images of sub-arcsec resolution with the Australia Telescope Compact Array (ATCA) (\citealp{gae97,man05,ng08,pot09,zan13}).
ATCA observations at 94 GHz \citep{lak12} have been followed by observations from 100 GHz up to 680 GHz with the Atacama Large Millimeter/submillimeter Array (ALMA;  \citealp{kam13,ind14}).

The ongoing shock expansion has been monitored at 9 GHz since 1992 (\citealp{gae97,gae07}). Shock velocities of $\sim$4000 km s$^{-1}$ have been measured between day 4000 and 7000 \citep{ng08}, while signs of a slower expansion have been tentatively detected after day $\sim$7000, as the shock has likely propagated past the high-density CSM in the ER \citep{ng13}.
Similar evidence of slower shock expansion since day $\sim$6000 has been found in X-ray data (\citealp{par05,par06, rac09}) as well as in  infrared (IR) data \citep{bou06}.

Since the early super-resolved images at 9 GHz (\citealp{gae97}), a limb-brightened shell morphology has been characteristic of the remnant. 
The radio emission, over the years, has become more similar to an elliptical ring rather than the original truncated-shell torus \citep{ng13}.
The radio remnant has shown a consistent  east-west asymmetry peaking on the eastern lobe, which has been associated with higher expansion velocities of the eastbound shocks \citep{zan13}. The asymmetry degree appears to have changed with the shock expansion, as images at 9 GHz exhibit a decreasing trend in the east-west asymmetry since day $\sim$7000 \citep{ng13}. High-resolution observations at 1.4--1.6 GHz \citep{ng11} via Very Long Baseline Interferometry (VLBI) with the Australian Large Baseline Array (LBA), have highlighted the presence of small-scale structures in the brightest regions in both lobes.

The relation between the radio emission and the synchrotron spectral indices, $\alpha$ ($S_{\nu}\propto\nu^{\alpha}$), has been investigated via both flux monitoring (\citealp{man02,zan10}) and imaging observations (\citealp{pot09,lak12,zan13}) with the ATCA. 
The progressive flattening of the radio spectrum derived from 843 MHz to 8.6 GHz at least since day 5000,  coupled with the $e$-folding rate of the radio emission, has pointed to an increasing production of non-thermal electrons and cosmic rays (CR) by the shock front \citep{zan10}. 
On the other hand, the association of steeper spectral indices with the brightest eastern sites implies a higher injection efficiency on the eastern side of the SNR \citep{zan13}. Flatter spectral indices in the center of the remnant have been tentatively identified from low-resolution two-frequency spectral maps \citep{pot09,lak12}, while at 18--44 GHz the central and center-north regions have $-0.5\lesssim\alpha\lesssim-0.3$ \citep{zan13}.
With ALMA, the spectral energy distribution (SED) of the remnant has been mapped where the non-thermal and thermal components of the emission overlap, 
identifying cold dust in the SNR interior \citep{ind14},  which accommodates $0.4-0.7\, M_{\sun}$ of the dust mass discovered with the {\it Herschel Space Observatory} ({\it Herschel}; \citealt{mat11}) in the ejecta.

This paper combines the results presented by \citet{ind14} with a comprehensive morphological and spectral analysis of SNR 1987A based on both ATCA and ALMA data.
In \S~\ref{Obs}, we present the ALMA Cycle 0 super-resolved images before and after subtraction, in the Fourier plane, of the synchrotron and dust  components (\S~\ref{morph}). In \S~\ref{Asym}, we assess the remnant asymmetry from 44 to 345 GHz. In \S~\ref{SED}, we update the SED derived by \citet{ind14}, while, in \S~\ref{SI}, we investigate the spectral index variations in the SNR across the transition from radio to far infrared (FIR). 
In \S~\ref{PWN--NS}, we discuss possible particle flux injection by a pulsar situated  in the inner regions of the remnant.

%
%
\begin{deluxetable*}{lcccc}[!thp]
\tablecaption{ALMA Observing Parameters \label{tab01}}
\tablewidth{0pt}
\tablecolumns{5}
\tablehead{\colhead{Parameter} 		& 
		 \colhead{102 GHz} 		& 
 		 \colhead{213 GHz} 		& 
		 \colhead{345 GHz} 		& 
		 \colhead{672 GHz} 		\\
		 \colhead{} 				& 
		 \colhead{(Band 3, {\it B3})}	& 
		 \colhead{(Band 6, {\it B6})}	& 
		 \colhead{(Band 7, {\it B7})}	& 
		 \colhead{(Band 9, {\it B9})}	 		 
}
\startdata
Date (2012)		         							& Apr 5, 6 		
											& Jul 15 \& Aug 10			
											& Jul 14 \& Aug 24			
											& Aug 25 \& Nov 5\\     	
Day since explosion								& 9174 				
											& 9287   					
											& 9294				
											& 9351\\  

Frequency bands\tablenotemark{$\ast$}  (GHz)			& 100.093--101.949
											& 213.506--213.597  
											& 336.979--340.917   
											& 661.992--665.992\\
											& 102.051--103.907 
											& 
											& 349.010--352.963  
											& 678.008--682.008\\

Center frequency (GHz) 							& 101.918 		
											& 213.146 				
											& 345.364		 	
											& 672.165\\
Channel width (MHz) 								& 4.883 
											& 4.883
											& 31.250
											& 15.625\\											
Max baselines  ($u,v$) (k$\lambda$) 					&150, 120
											&180, 260
											&400, 400
											&700, 700\\
No. of antennas 									& 14--18 			& 14--23			&  28 				 &  19--25\\
Total observing time (hr) 							& 0.83 			& 1.03 			&  0.62 			 &  3.40
\enddata
\tablenotetext{$\ast$}{In B3 and B6, the frequency range is selected to avoid CO and SiO emission \citep{kam13}.}
\vspace{0.0mm}
\end{deluxetable*}

%
%
\begin{deluxetable*}{lcccccc}[!thp]
\tablecaption{Image Parameters \label{tab02}}
\tablewidth{0mm}
\tablecolumns{7}
\tablehead{\multicolumn{1}{l}{Image} 					& 
		 \colhead{$S_{\nu}$}\tablenotemark{${(a)}$} 		& 
 		 \colhead{SR Beam}\tablenotemark{${(b)}$}		& 
		 \colhead{DL Beam}\tablenotemark{${(c)}$}		& 
		 \colhead{PA} 							& 
		 \colhead{Rms noise} 						& 
		 \colhead{Dynamic range} 					\\
		 \multicolumn{1}{l}{(GHz)} 					&
		 \colhead{(mJy)} 							& 
 		 \colhead{($''$)}							& 
		 \colhead{($''$)}							&
		 \colhead{($^{\circ}$)	} 					&
		 \colhead{(mJy/beam)} 					&		  
		 \colhead{} 				
}
\startdata
$\;\;$94\tablenotemark{$^{(d)}$}			& $24.2\pm3.9$			& $0.7$			& \nodata				& \nodata				& $0.085$			& $137$\\ 
102\tablenotemark{$^{(e)}$}				& $23.1\pm3.1$			& $0.8$ 			& $1.74\times1.25$		& $5.7$				& $0.033$			& $285$\\ 
213								& $19.7\pm1.6$			& $0.6$ 			& $1.16\times0.74$		& $-68.5$				& $0.034$			& $75$\\
345								& $16.7\pm1.5$			& $0.3$			& $0.65\times0.48$		& $-40.2$				& $0.023$			& $121$\\
672								& $52.8\pm14.2$			& $0.3$ 			& $0.34\times0.28$		& $68.5$				& $1.219$			& $28$\\
\cmidrule(lr){1-7}	
$\;\;$94$-I_{\rm B9}$\tablenotemark{$^{(f)}$}  & $23.2\pm3.9$			&  $0.7$			& $0.78\times0.63$		& $15.4$				& $0.094$			& $81$\\
102$-I_{\rm B9}$						& $19.4\pm3.2$			&  $0.8$ 			& $1.66\times1.19$		& $6.3$				& $0.034$			& $143$\\
213$-I_{\rm B9}$						& $16.9\pm1.9$			&  $0.6$			& $1.16\times0.74$		& $-68.5$				& $0.029$			& $64$\\
345$-I_{\rm B9}$						& $11.5\pm1.9$			&  $0.3$			& $0.65\times0.48$		& $-44.0$				& $0.031$			& $43$\\
672$-I_{\rm B9}$						& $-1.0\pm14.3$ 			&  $0.3$ 			& $0.34\times0.28$		& $68.5$				& $1.193$			& $2$\\ 

\cmidrule(lr){1-7}
$\;\;$94$-I_{44}$\tablenotemark{$^{(g)}$}	& $0.9\pm3.9$			& $0.7$			& $ 0.78\times0.63$		& $15.4$				& $0.095$			& $60$\\
102$-I_{44}$						& $3.5\pm3.1$			& $0.8$			& $1.48\times0.96$		& $10.1$				& $0.027$			& $81$\\
213$-I_{44}$						& $2.9\pm1.7$			& $0.6$ 			& $1.16\times0.74$		& $-68.5$				& $0.024$			& $26$\\ 
345$-I_{44}$						& $5.8\pm1.6$			& $0.3$			& $0.65\times0.48$		& $-44.0$				& $0.014$			& $90$\\
672$-I_{44}$						& $47.4\pm14.8$			& $0.3$ 			& $0.57\times0.55$		& $78.6$				& $0.816$			& $24$
\enddata
\vspace{0.5mm}
\tablenotetext{$a$}{All flux densities are derived from the diffraction-limited images. The errors are derived from the flux calibration uncertainty combined with the uncertainties in the image subtraction.}.
\tablenotetext{$b$}{Circular beam used for super-resolution (SR).}
\tablenotetext{$c$}{Beam associated with the diffraction-limited (DL) image.} 
\tablenotetext{$d$}{The flux density is scaled to day 9280 via  exponential fitting parameters derived for ATCA flux densities from day 8000, as measured at 8.6 and 9 GHz (Zanardo et al., in preparation). All other image parameters are as from \citet{lak12}.}
\tablenotetext{$e$}{Images are shown in Figures~\ref{fig:df} and \ref{fig:super}.}
\tablenotetext{$f$}{Images obtained by subtracting the model flux density at 672 GHz ($I_{\rm B9}$) scaled to fit the central emission. See central column of  Figures~\ref{fig:df} and \ref{fig:super}.}
\tablenotetext{$g$}{Images obtained by subtracting the model flux density at 44 GHz ($I_{44}$) scaled to fit the toroidal emission. See left column of  Figures~\ref{fig:df} and \ref{fig:super}.
\\ \\}
\end{deluxetable*}
\vspace{1.0mm}

\section{Observations and Analysis}
\label{Obs}

The ATCA and ALMA observations used in this study were performed in 2011 and 2012. ATCA observations at 44 and 94 GHz are detailed in \citet{zan13} and \citet{lak12}, respectively.
ALMA observations  were made in 2012 (Cycle 0) from April to November, over four frequency bands: Band 3 (B3, 84--116 GHz, $\lambda$ 3 mm), Band 6 (B6, 211--275 GHz, $\lambda$ 1.3 mm), Band 7 (B7, 275--373 GHz, $\lambda$ 850 $\mu$m) and Band 9 (B9, 602--720 GHz, $\lambda$ 450 $\mu$m). Each band was split over dual 2-GHz-wide sidebands, with minimum baselines of 17 m (B9) to maximum baselines of 400 m (B3). 
All observations used quasars J0538-440 and J0637-752 as bandpass and phase calibrators, respectively. 
Callisto was observed as an absolute flux calibrator in B3 and B6, while Ceres was used in B7 and B9 (see also \citealp{kam13}).
It is noted that, while ALMA is designed to yield data with flux density calibration uncertainty as low as $\sim$1\%, in Cycle 0 this uncertainty is estimated at $\sim$5\% at all frequencies.
Relevant observational parameters are listed in  Table~\ref{tab01} (see also Table 1 in \citealt{ind14}).

Each dataset was calibrated with the {\sc casa}\footnote[1]{http://casa.nrao.edu/} package, then exported in {\sc miriad}\footnote[2]{http://www.atnf.csiro.au/computing/software/miriad/} for 
imaging.  After {\sc clean}-ing  \citep{hog74}, both phase and amplitude self-calibration were applied in B3 over a 2-minute solution interval, while only phase calibration was applied in B6 and B7.
No self-calibration was performed in B9. 
As in \citet{zan13}, we note that since the self-calibration technique removes position information, each image was compared with that
prior to self-calibration and, in case of positional changes, the self-calibrated images were shifted. Further adjustments were made in the comparison with the ATCA observations at 44 GHz, based on prominent features on the eastern lobe and location of the  remnant center. As from \citet{zan13}, the 44 GHz image was aligned with VLBI observations of the SNR (Zanardo et al. in preparation).
Adding in quadrature these positional uncertainties  and the accuracy of the  LBA VLBI frame, the errors in the final image position are estimated at $\sim$60 mas.

%
%
\begin{figure*}[htp]
\begin{center}
\vspace{-2mm}
\advance\leftskip-4mm
\includegraphics[trim=0mm 22.0mm 10.0mm 22mm, clip=true,width=182mm, angle=0]{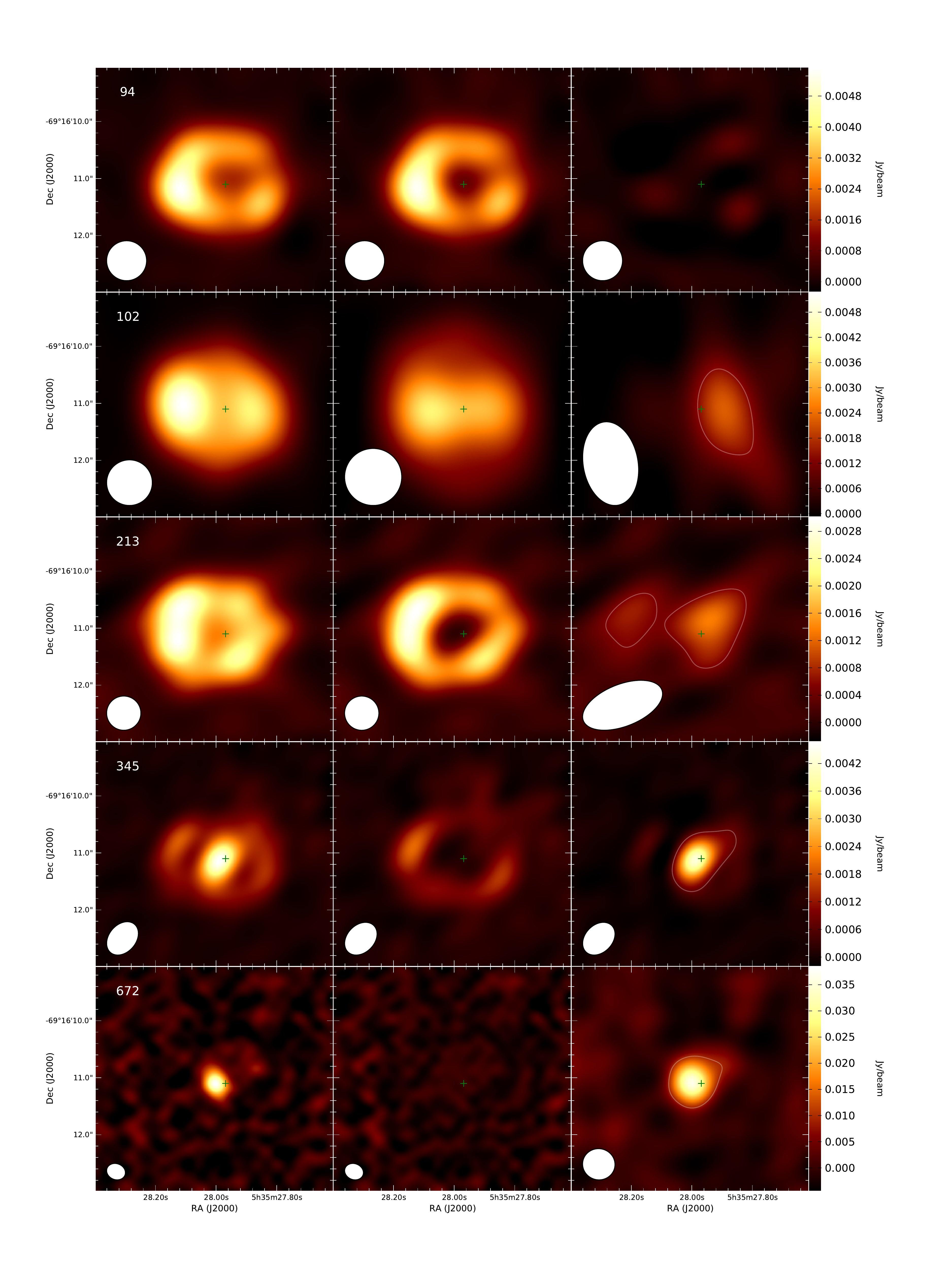}
\caption{
{\it Top to bottom -- Left column:} Stokes $I$ continuum images of SNR 1987A at 94 \citep{lak12}, 102, 213, 345 and 672 GHz.  Images from 102 to 672 GHz are made from ALMA observations (Cycle 0) performed from 2012 April 5 to November 5 (see Table~\ref{tab01}). {\it center column:} Images obtained by subtracting the model flux density at 672 GHz (Band 9) scaled to fit the central emission. {\it Right column:} Images obtained by subtracting a scaled model flux density at 44 GHz \citep{zan13},  with $3\,\sigma$ flux density contours highlighted ({\it white}).
The angular resolution is shown in the bottom left corner.
The green cross indicates the
 VLBI position of SN 1987A as determined by \citet{rey95} [RA $05^{\rm h}\;35^{\rm m}\;27\fs968$, Dec $-69^{\circ}\;16'\;11\farcs09$ (J2000)].}
\label{fig:df}
\end{center}
\end{figure*}

%
%
\begin{figure*}[htp]
\begin{center}
\vspace{-2mm}
\advance\leftskip-4mm
\includegraphics[trim=0mm 22.0mm 10.0mm 22mm, clip=true,width=182mm, angle=0]{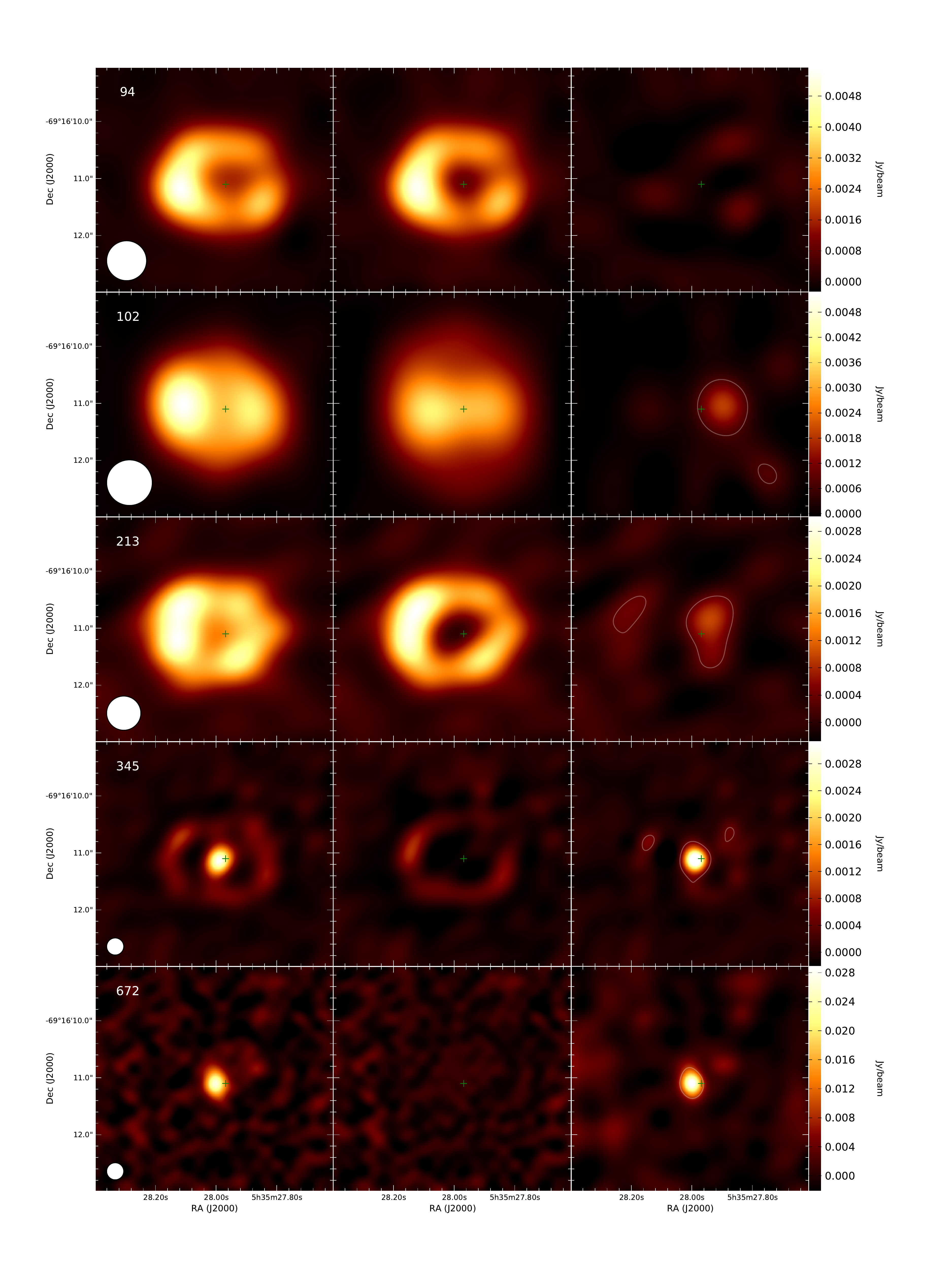}
\caption{The panel layout is identical to that in Figure~\ref{fig:df}, but all Stokes $I$ continuum subtracted images are super-resolved. The circular beam used for the super-resolved ALMA images is $0\farcs8$ at 102 GHz (Band 3), $0\farcs6$ at 213 GHz (Band 6), $0\farcs3$ at 345 (Band 7) and 672 GHz (Band 9), and is plotted in the lower left corner.  The 94 GHz images are restored with a $0\farcs7$ circular  beam. For images on the right column, $3\,\sigma$ flux density contours are highlighted ({\it white}). The green cross indicates the VLBI position of SN 1987A as determined by \citet{rey95}.}
\label{fig:super}
\end{center}
\end{figure*}

%
%
\begin{figure*}[!htbp]	 
\begin{center}
\vspace{5mm}
\advance\leftskip4mm
\includegraphics[trim=0mm 0.0mm 0.0mm 1.0mm, clip=true,width=128mm, angle=0]{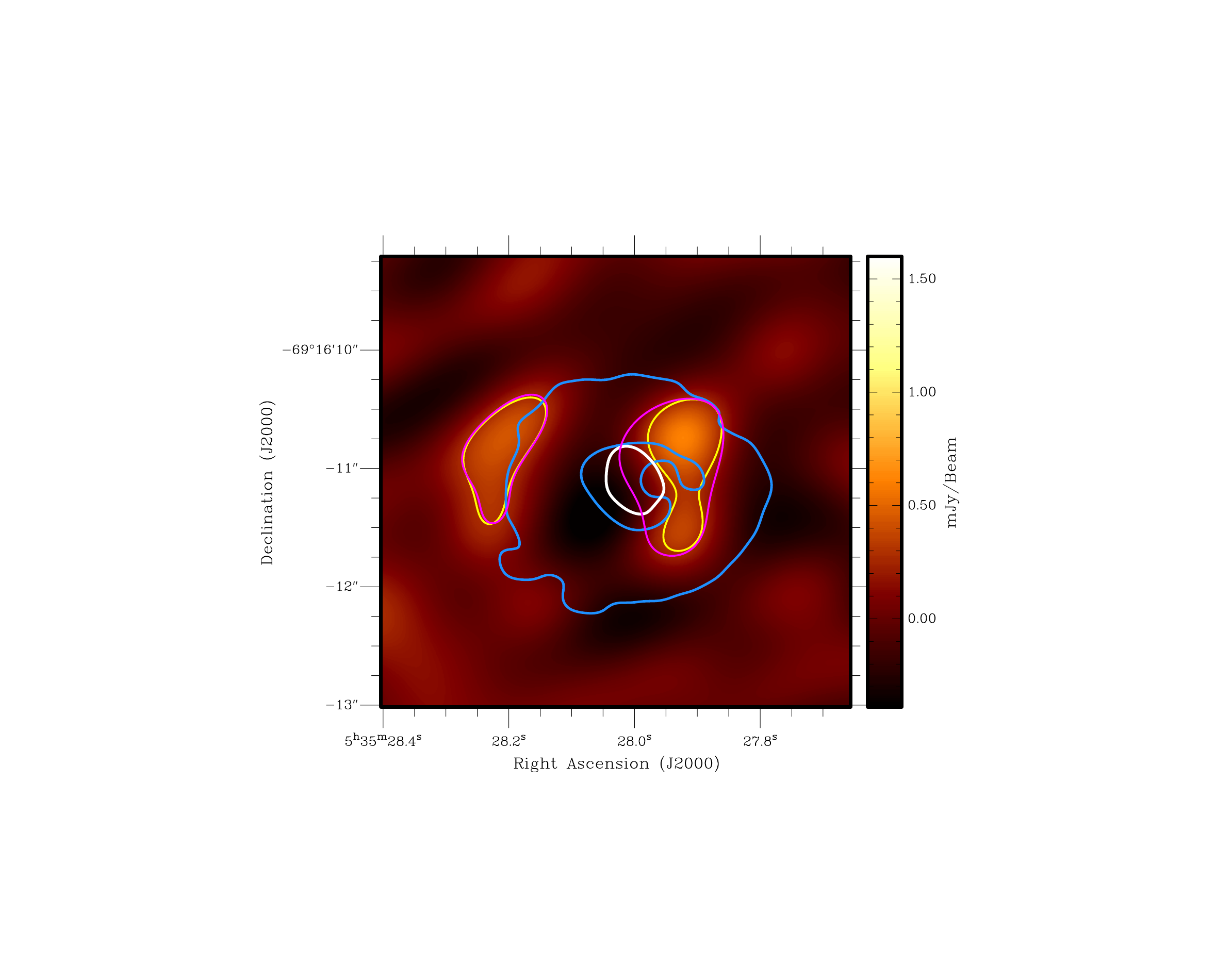}
\caption{
Stokes $I$ continuum image of SNR 1987A at 213 GHz as obtained after the dual subtraction of the model flux densities at 44  and 672 GHz (see  \S~\ref{Obs}). 
The image is super-resolved with a 0\farcs6 circular beam, and the yellow contour highlights the $3\,\sigma$ flux density levels.  The off-source rms noise is 0.14 mJy beam$^{-1}$ (S/N $\sim6\,\sigma$).
For comparison, the 213 GHz image obtained after the single subtraction of the model flux density at 44 GHz and similarly super-resolved with a 0\farcs6 circular beam (see Figure~\ref{fig:super}), is outlined via the $3\,\sigma$ flux density levels (magenta contours).
The model images, i.e. the Stokes $I$ images at 44 GHz (\citealp{zan13}) and at 672 GHz, are outlined by blue and white contours, respectively, at the $5\,\sigma$ flux density levels. Both model images are resolved with a 0\farcs3 circular beam. \\}
\label{fig:B6-44-B9}
\end{center}
\end{figure*} 
%
%

%
%
\begin{figure*}[!htbp]	 
\begin{center}
\vspace{0mm}
\advance\leftskip4mm
\includegraphics[trim=0mm 0.0mm 0.0mm 0mm, clip=true,width=138mm, angle=0]{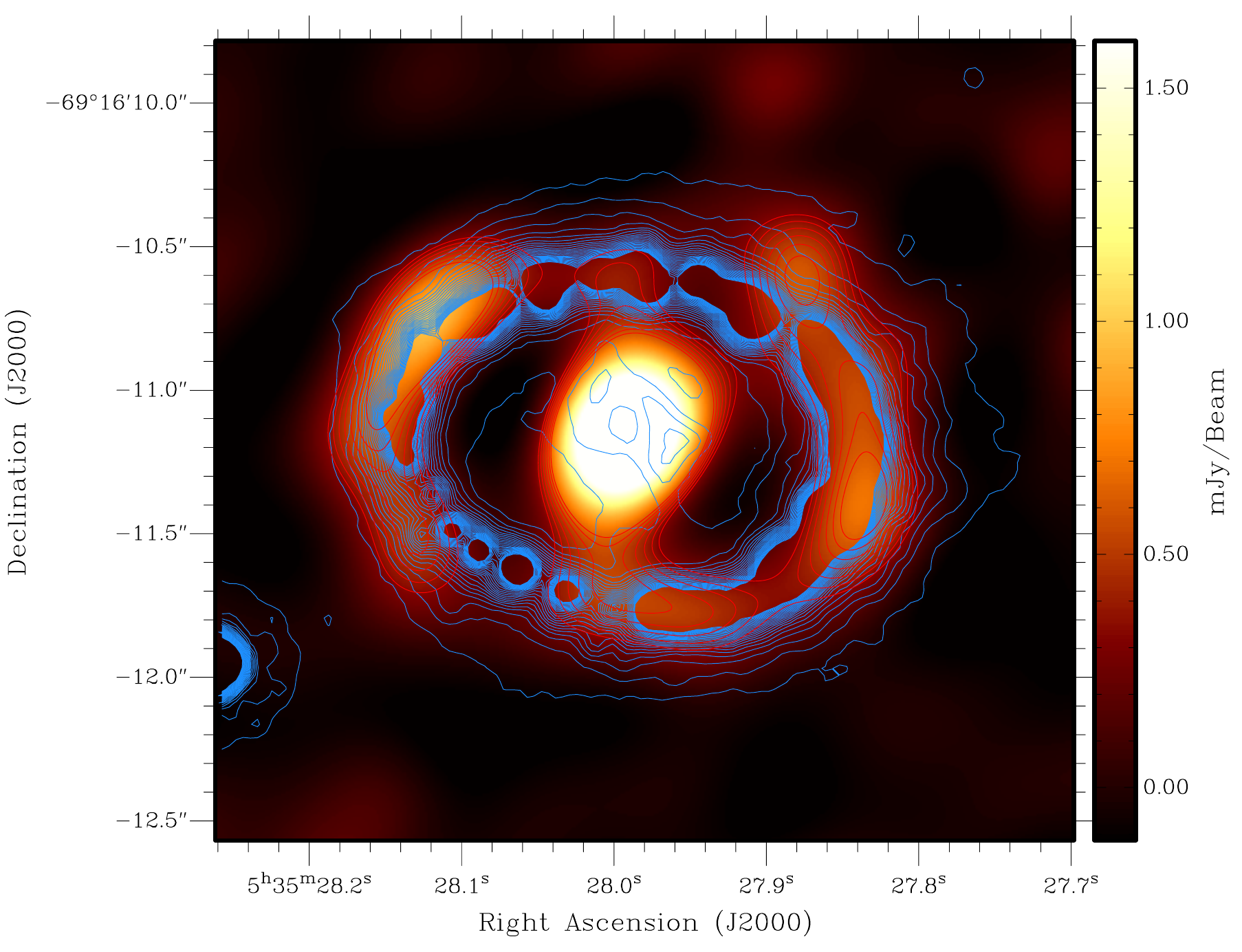}
\caption{Overlay of the HST image of SNR 1987A (light blue contours; \citep{lar11}) on the 345 GHz image produced from ALMA observations performed in 2012 June and August (brown--yellow color scale and red contours). As from Figure~\ref{fig:super}, the ALMA image is super-resolved with a $0\farcs3$ circular beam.\\ \\}
\label{fig:B7_Hbl}
\end{center}
\end{figure*} 

\vspace{0.0mm}
Deconvolution was carried out via the maximum entropy method (MEM) \citep{gul78} in B3, B6 and B7. A weighting parameter of robust = 0.5  \citep{bri95} was used in all bands. The resultant diffraction-limited images, which have central frequency at 102 GHz in B3, 213 GHz in B6, 345 GHz in B7 and 672 GHz in B9, were then super-resolved with a circular beam of $0\farcs8$ in B3, $0\farcs6$  in B6, and $0\farcs3$ in B7 and B9. The   diffraction limited and super-resolved images are shown in the first column of Figures~\ref{fig:df} and~\ref{fig:super}, below the ATCA image at 94 GHz  \citep{lak12}. Integrated Stokes $I$ flux densities, dynamic range and related rms are given in Table ~\ref{tab02}.  

To decouple the non-thermal emission from that originating from dust, the synchrotron component, as resolved with ATCA at 44 GHz \citep{zan13}, and the dust component, as imaged with ALMA at 672 GHz (B9) (\citealp{ind14}), were separately subtracted from the datasets at 94, 102, 213, 345 and 672 GHz. All subtractions were performed in the Fourier plane, via {\sc miriad} task {\tt uvmodel}, where the model flux density at 44 GHz was scaled to fit the SNR emission over the ER ($I_{44}$), while the B9 model flux density was scaled to fit the emission localised in the central region of the remnant ($I_{\rm B9}$). Scaling of the 44 GHz model was tuned by minimizing the  flux density difference on the brighter eastern lobe, without over-subtracting in other regions of the remnant. To separate the emission in the SNR center, the central flux was firstly estimated by fitting a gaussian model via {\sc miriad} task {\tt uvfit}. The image model at 672 GHz was then scaled to match the flux of the gaussian model. The scaling factor was further tuned to minimize over-subtraction. 

Super-Nyquist sampling was applied in all images, using a pixel size of 8 mas to avoid artefacts when sources are not at pixel centers. 
Deconvolution via MEM was carried out on the residual images obtained from the subtraction of $I_{\rm B9}$, while standard {\sc CLEAN}ing was applied to the residuals obtained from the $I_{44}$ subtraction. 
All diffraction-limited subtracted images are shown in the central and right columns of Figure~\ref{fig:df}, while Figure \ref{fig:super} shows the residuals after super-resolution with the circular  beam used for the original images.  
All image parameters are given in Table ~\ref{tab02}.

The flux densities were determined by integrating within polygons enclosing the SNR emission. 
Uncertainties in the flux densities include uncertainties in the image fitting/scaling process combined with the uncertainty in the flux density calibration.
We note that the residual images from the subtraction of both models at 44 and 672 GHz were not considered, since the error attached to the double subtraction exceeds the total integrated flux density. 
Only at 213 GHz  the error--flux margin is minor, as the diffraction-limited image obtained after the dual subtraction has integrated flux density of $2.1\pm1.9$ mJy (Figure~\ref{fig:B6-44-B9}).

%
%
\begin{figure*}[htp] 
\begin{center}
\vspace{-0mm}
\advance\leftskip4mm
\includegraphics[trim=0mm 0.0mm 0.0mm 0mm, clip=true,width=146mm, angle=0]{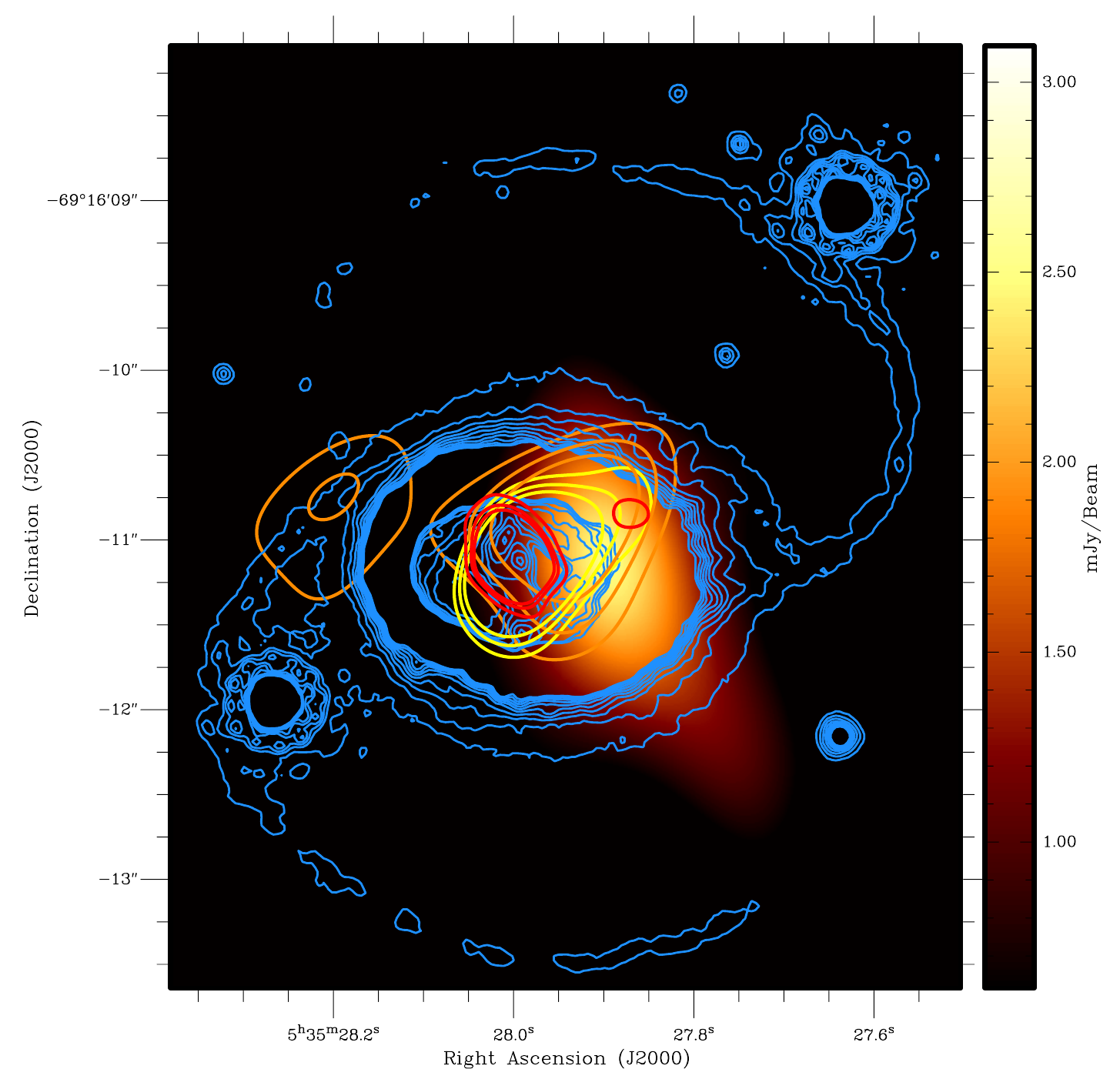}
\caption{
Comparison of the diffraction-limited images of SNR 1987A at 102, 213 and 345 GHz, as obtained after subtraction of the scaled model flux density at 44 GHz (see Figure~\ref{fig:df}, right column), with the optical image of the remnant. In detail, the residual image at 102 GHz (brown--yellow color scale) is overlaid with the contours outlining the residual images at 213 GHz ({\it orange}) and at 345 GHz ({\it yellow}). To locate the main sites associated with dust emission, the diffraction-limited image at 672 GHz is also outlined ({\it red}) (as from Figure~\ref{fig:df}, left column).  Images at 213, 345 and 672 GHz are overlaid with contours at 3, 4 and 5\,$\sigma$ flux density levels.  The angular beams associated with each image, as shown in Figure~\ref{fig:df},  are listed in Table~\ref{tab02}.
The {\it HST} image \citep{lar11} is outlined via contours ({\it blue}) that highlight the structure of the outer rings, the equatorial ring and the ejecta.\\ 
}
\label{fig:B3679_Hbl}
\end{center}
\end{figure*} 

\section{Morphology}
\label{morph}

ALMA observations of SNR 1987A capture both the remnant emission from the ER and that from the SNR interior, where the dense ejecta sit (\citealp{kam13,ind14}). 

While the image at 102 GHz barely resolves the two-lobe distribution of the emission, the images at 213 and 345 GHz clearly show the ringlike emission morphology, localised around the ER  (see Figures~ \ref{fig:df} and \ref{fig:super}).
It is understood that the radio emission over the ER is primarily synchrotron emission, generated by the interaction of the SN shock with the dense CSM near the equatorial plane, which results in a magnetic-field discontinuity where particles are accelerated (e.g. \citealp{zan10,zan13}). 
As shown in early models of Type II SNRs \citep{che82},
the region of interaction between the SN blast and the CSM consists of a double-shock structure, with a forward shock, where ambient gas is compressed and heated, and a reverse shock, where the ejecta are decelerated. Between the two shocks, the reflected shocks, due to the forward shock colliding with the dense ER \citep{bor97}, propagate inward (\citealp{zhe09,zhe10}).
The reverse shock at first expanded outwards, behind the forward shock, but might have been inward-moving since day $\sim7000$.
As discussed by \citet{ng13}, the ring synchrotron emission is currently localised between the forward and reverse shocks, and likely has components from both the ER and high-latitude material above the equatorial plane. Truncated-shell torus models of the remnant geometry at 9 GHz indicate that the half-opening angle has been decreasing  since day $\sim$7000, and is estimated at $\sim27^{\circ}$ at day 9568 \citep{ng13}. 

At 345 GHz the SNR interior is brighter, while at 672 GHz the emission is predominantly localised in the central region of the remnant. 
Since the emission from this region rises steeply with frequency, it has been identified with thermal dust emission \citep{ind14},
as dust grains, probably heated by $^{44}$Ti decay and X-ray emission from the reverse shock \citep{lar13}, emit strongly in the FIR regime.
The central emission, visible both in B7 and B9,  appears to extend over the inner optical ejecta (see Figure~\ref{fig:B7_Hbl}). 
In particular,  as noted by \citet{ind14}, this inner emission shows a north--south elongation, which in B9 can be identified  between PA $\sim20^{\circ}$ and PA $\sim30^{\circ}$,
similar to that seen with {\it HST} (\citealp{lar11,lar13}).
From both Figures~\ref{fig:df} and \ref{fig:super}, it can also be noticed that the SNR emission at 672 GHz includes possible emission located to the NW  (see contour overlays in Figure~\ref{fig:B3679_Hbl}).
This NW feature  has signal-to-noise ratio (S/N) of $3.7\,\sigma$, and integrated flux density of 6.8$\pm$0.5 mJy, i.e. $\sim$10\% of the total integrated flux density at 672 GHz.

\subsection{Subtracted Images}
 \label{Subtract}

To identify the origin of the emission in the  $I_{44}-$subtracted images, with respect to the structure of the remnant as seen with {\it HST} \citep{lar11}, the diffraction-limited  residuals at 102, 213 and 345 GHz are superimposed in Figure~\ref{fig:B3679_Hbl}. 
It can be seen that the residual at 102 GHz, 
characterised by  $\left. {\rm S/N} \right |_{102} =6.9\,\sigma$,  
is mainly located on the western lobe, west of the 
VLBI position of SN 1987A, as determined by \citet{rey95} [RA $05^{\rm h}\;35^{\rm m}\;27\fs968$, Dec $-69^{\circ}\;16'\;11\farcs09$ (J2000)].
We note that, since the residual emission at 102 GHz has a constant flux per synthesized beam,  the S/N ratio is independent on the beam size. 
The emission at 213 GHz (orange contours in Figure~\ref{fig:B3679_Hbl}), 
with $\left. {\rm S/N} \right |_{213} =7.7\,\sigma$,
peaks NW of the SN position, while  fainter emission
($\left. {\rm S/N}\right |_{213}^{E}   =4.2\,\sigma$)
may extend NE.  
The residual emission at both 102 and 213 GHz
is above noise levels and, thus, unlikely to be the result of image artefacts.
Given the brightening of the emission from the dust in the central region of the SNR, the residual images at 345 and 672 GHz have higher S/N. 
In particular, at 345 GHz (yellow contours in Figure~\ref{fig:B3679_Hbl}) $\left. {\rm S/N} \right |_{345} =17.4\,\sigma$, while, similar to the morphology at 213 GHz, the residual emission extends westwards and elongates NW, with a much fainter spot on the north-eastern section of the ER ($\left. {\rm S/N}\right |_{345}^{E}   =2.4\,\sigma$).
A  westward-elongated  morphology is present in the image at 672 GHz ($\left. {\rm S/N} \right |_{672}=12.5\,\sigma$).

The  $I_{\rm B9}-$subtracted images emphasize 
the ringlike morphology of the synchrotron emission that mainly originates  near the ER (see \S~\ref{morph}) and, thus,
the asymmetry  of this emission, which is discussed in \S~\ref{Asym}.
These residuals also highlight the presence of  NW emission at 672 GHz, i.e. outside the inner SNR (see red contours in Figure~\ref{fig:B3679_Hbl}). In fact, at 94, 213, and 345 GHz, it can be seen that the discontinuity in the NW sector of the ER, between PA $\sim290^{\circ}$ and PA $\sim300^{\circ}$, becomes more prominent after subtraction of the B9 model.
Some extended emission north and NW of the ER emerges from noise at 345 GHz.

\section{Asymmetry}
\label{Asym}

%
%
\begin{figure*}[htp]
\begin{center}
\vspace{5mm}
\advance\leftskip-3.0mm
\includegraphics[trim=0mm 0.0mm 0.0mm 0mm, clip=true,width=186mm, angle=0]{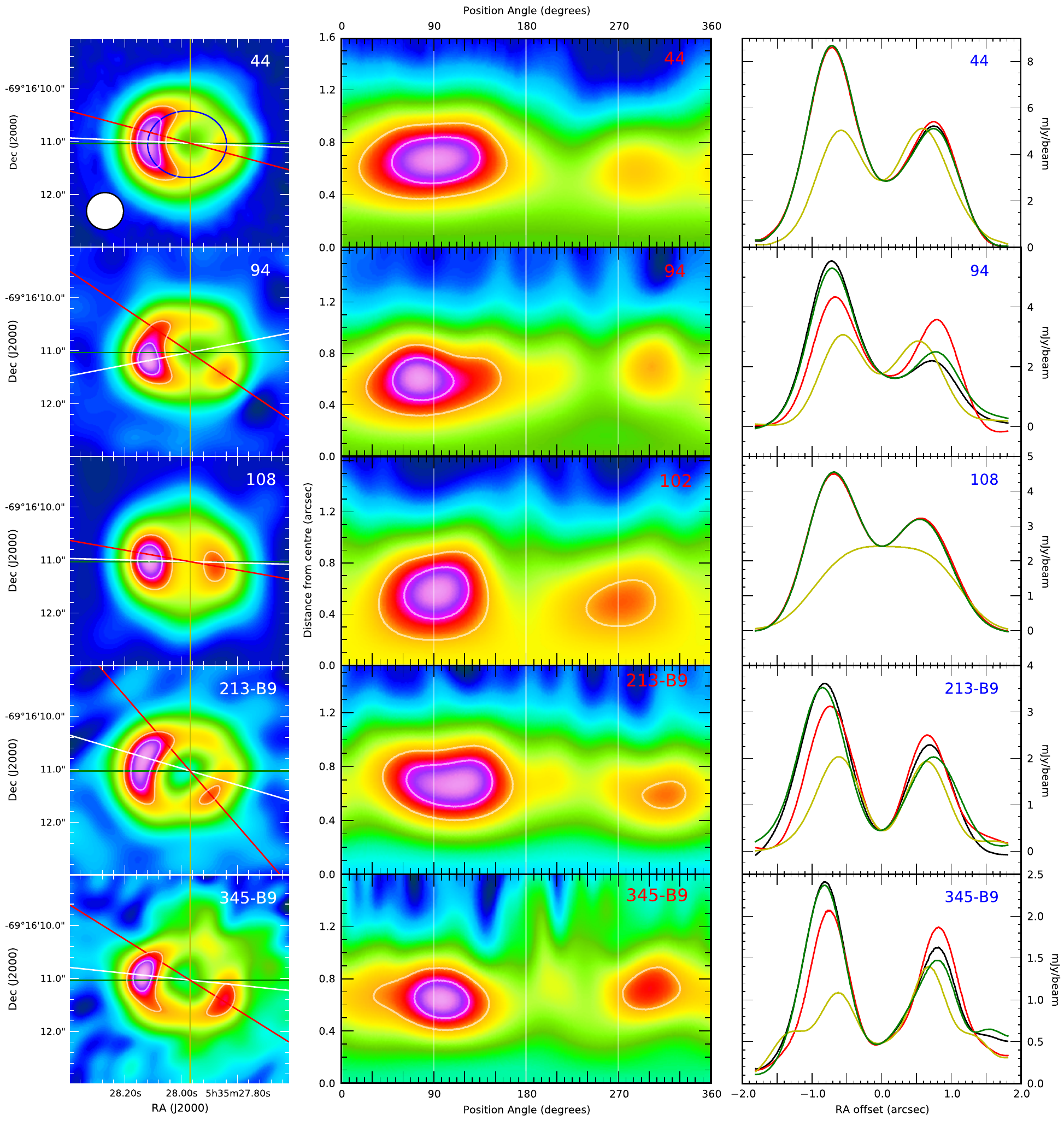}
\caption{
{\it Top to bottom - Left column}: Stokes $I$ continuum images of SNR 1987A at 44, 94, 102, 213, and 345 GHz. The 213 ({\it 213--B9}) and 345 GHz ({\it 345--B9}) images are derived after subtraction of the scaled model flux density at 672 GHz (Band 9, {\it B9}). All images are resolved with a $0\farcs7$ circular beam (as shown in the bottom left corner of the 44 GHz image). 
{\it Center column:} The images on the left column are converted to polar coordinates to visualize asymmetries in the two-dimensional radial distribution. The position angle and the projected radial distance from the geometrical center of the remnant, are the new coordinates. Each image is projected along concentric ellipses (e.g. {\it blue} ellipse overlaid on the 44 GHz image). The emission intensity is conserved in the conversion. Contours at 65\% and 85\% levels of the emission are shown on all images ({\it white}).
{\it Right column:} Radial slices through each image at four position angles (as indicated on the images on the left column), which correspond to  $0^{\circ}$ ({\it light green}), $90^{\circ}$ ({\it dark green}), emission peaks on the eastern lobe ({\it white/black}) and on the western lobe ({\it red}). The radial slices are plotted against the RA offset from the 
 VLBI position of SN 1987A as determined by \citet{rey95} [RA $05^{\rm h}\;35^{\rm m}\;27\fs968$, Dec $-69^{\circ}\;16'\;11\farcs09$ (J2000)].
}
\label{fig:Asym}
\end{center}
\end{figure*}  

The east-west asymmetry of the synchrotron emission, primarily associated with the emission morphology over the ER, is investigated from 44 to 345 GHz,  as shown in Figure~\ref{fig:Asym}, where all images are restored with a $0\farcs7$ circular beam. At  213 and 345 GHz the central dust emission is subtracted, via a scaled model of the flux density as resolved in B9 (see \S~\ref{Obs}). 

\begin{figure}[!t] 
\begin{center}
\vspace{0.0mm}
\advance\leftskip-3mm
\includegraphics[trim=6.5mm 5.0mm 10.0mm 10.0mm, clip=true,width=89mm, angle=0]{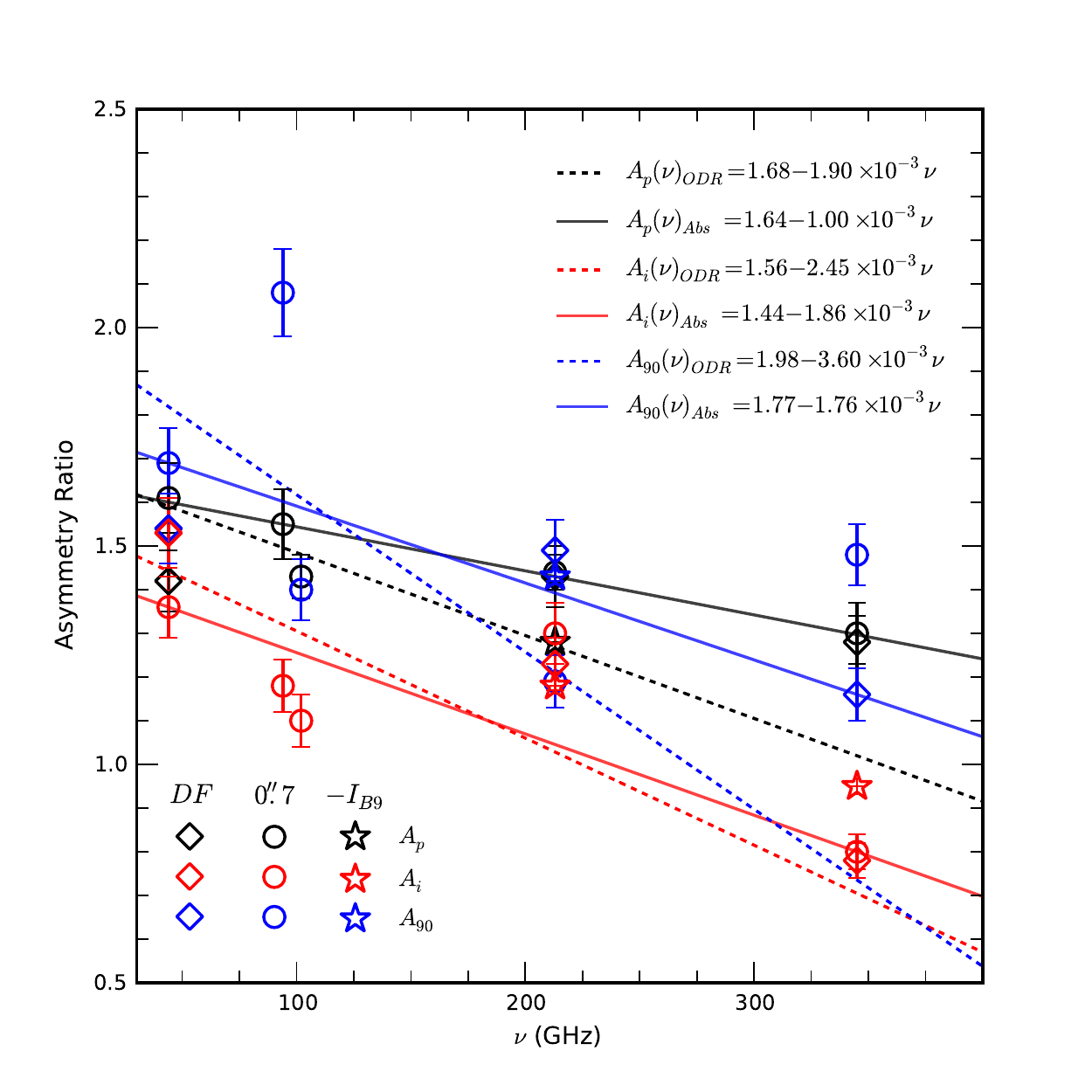}
\caption{
Asymmetry ratios of Stokes $I$ images of SNR 1987A from 44 to 345 GHz. With reference to Table~\ref{tab_asym}, the east-west asymmetry ratios estimated via three different methods are indicated as: $A_{p}$ (black symbols);   		
$A_{i}$ (red symbols);  		
$A_{90}$ (blue symbols).  	
Asymmetry ratios are derived from the diffraction-limited images ({\it diamonds}) (see Figure~\ref{fig:df}), from the images restored with a $0\farcs7$ circular beam ({\it circles}) (see Figure~\ref{fig:Asym}); and from images obtained after the subtraction of the dust emission component ({\it stars}) (see  Figure~\ref{fig:Asym}). Tentative linear fits are shown for $A_{p}$  (black fits), $A_{i}$ (red fits) and $A_{90}$ (blue fits), based on all ratios derived via each method.  The 
orthogonal distance regression ($A_{ODR}$) and the robust fitting from the square root of absolute residuals ($A_{Abs}$), are tested for linear interpolation.
The combined fits for  the $A_{p}$ and $A_{i}$ ratios yield an average $A(\nu)\sim A(\nu_{0}) -1.9\pm0.8\times10^{-3} \nu$, where $\nu$ is expressed in GHz and $A(\nu_{0}=30)=1.57\pm0.13$. \\
}
\label{fig:Asym_ratios}
\end{center}
\end{figure}

Considering radial sections crossing the SN site \citep{rey95}, the SNR asymmetry is firstly estimated as the ratio between the eastern peak of the radial slice crossing the maximum emission on the eastern lobe 
(see {\it black/white} profile in Figure~\ref{fig:Asym}), and the western peak of the radial slice crossing the maximum emission on the western lobe, 
(see {\it red} profiles in Figure~\ref{fig:Asym}). This ratio, $A_{p}$, emphasizes the hot spots on each side of the remnant. Alternatively, the ratio  $A_{i}$ between the total flux densities integrated over the eastern and western halves of the image, is derived by splitting the image at the RA associated with the SN site. 
For comparison with previous asymmetry estimates \citep[e.g.][]{zan13}, the ratio $A_{90}$ between the eastern and western peaks of the radial slice at PA $90^{\circ}$ (see {\it dark green} profile in Figure~\ref{fig:Asym}) is also derived. 
All values for $A_{p}$, $A_{i}$ and $A_{90}$, both for the images restored with a $0\farcs7$ circular beam (as in Figure~\ref{fig:Asym}) and the diffraction-limited images, are plotted in Figure ~\ref{fig:Asym_ratios} and listed in Table~\ref{tab_asym}.

From  Figure~\ref{fig:Asym}  it can be seen that the $A_{90}$ ratio does not fully capture the asymmetry changes with frequency, since the hot spot in the western lobe, which becomes brighter from 102 to 345 GHz, is located southwards of the 90$^{\circ}$ profile.
In Figure~\ref{fig:Asym_ratios}, the linear fits derived for both $A_{p}$ and $A_{i}$ ratios show a consistent decrease as frequencies reach the FIR. 
At 345 GHz, $A_{i}$ values indicate that the east-west asymmetry is reversed, thus matching the asymmetry trend seen in recent {\it HST} images \citep{lar11}, where the western side of the ring is markedly brighter. 
As discussed in \S~\ref{morph}, the morphology similarities between the optical image of the SNR and the super-resolved image at 345 GHz are evident (see Figure~\ref{fig:B7_Hbl}).

The change of the remnant's east-west asymmetry over time has been discussed by \citet{ng13}, as the result of a progressive flattening of the shock structure in the equatorial plane, due to the shock becoming engulfed in the dense UV-optical knots in the ER,  coupled with faster shocks in the east side of the remnant. 
While X-ray observations do not show significant difference between the NE and SW reverse shock velocities, although the NE sector is brighter \citep{fra13},  
faster eastbound outer shocks have been measured in the radio \citep{zan13} and point to an asymmetric explosion of a binary merger as SN progenitor (\citealp{mor07,mor09}). As the SN blast is gradually overtaking the ER, faster expanding shocks in the east would exit the ER earlier than in the west.

%
%
\begin{center}
\begin{deluxetable*}{lcccccc}[!t]
\tablecaption{Asymmetry Ratios \label{tab_asym}} 
\tablewidth{0mm}
\tablecolumns{7}
\tablehead{
\colhead{Image} 																					&
\multicolumn{2}{c}{$A_{p}\equiv\dfrac{S_{\nu}(E_{Max})} {S_{\nu}(W_{Max})}$ \tablenotemark{$^{(a)}$}} 					&  
\multicolumn{2}{c}{$A_{i}\equiv\slfrac{ \int\limits_E S_{\nu}}  {\int\limits_W S_{\nu}} $ \tablenotemark{$^{(b)}$}} 					& 	
\multicolumn{2}{c}{$A_{90}\equiv\left. S_{\nu} \left (\dfrac{E_{Max}} {W_{Max}} \right) \right |_{90^{\circ}}$\tablenotemark{$^{(c)}$}} 	\\					
}
\startdata
\multicolumn{1}{l}{GHz}						&  
\colhead{$0\farcs7$\tablenotemark{$^{(d)}$}} 			& 	
\colhead{DL \tablenotemark{$^{(e)}$}}				&
\colhead{$0\farcs7$} 							& 	
\colhead{DL}	  							&	
\colhead{$0\farcs7$} 							& 	
\colhead{DL}	  							\\						 
\cmidrule(lr){2-3}
\cmidrule(lr){4-5}		
\cmidrule(lr){6-7}\\

$\;\;$44							&	$1.61\pm0.08$	&	$1.42\pm0.05$ 	&	$1.36\pm0.02$  	&	$1.53\pm0.07$	&	$1.69\pm0.08$ 	&	$1.54\pm0.05$\\
$\;\;$94							&	$1.55\pm0.05$	&	\nodata 		&	$1.18\pm0.04$  	&	\nodata		&	$1.88\pm0.07$ 	&	\nodata\\
102								&	$1.44\pm0.04$	&	\nodata 		&	$1.10\pm0.04$  	&	\nodata		&	$1.40\pm0.07$ 	&	\nodata\\
213								&	$1.28\pm0.03$	&	$1.43\pm0.05$ 	&	$1.24\pm0.06$  	&	$1.23\pm0.02$	&	$1.43\pm0.02$ 	&	$1.49\pm0.05$\\	
213$-I_{\rm B9}$\tablenotemark{${(f)}$}		&	$1.44\pm0.03$	&	\nodata 		&	$1.30\pm0.02$  	&	\nodata		&	$1.19\pm0.03$ 	&	\nodata\\
345								&	\nodata		&	$1.28\pm0.03$ 	&	\nodata  		&	$0.78\pm0.04$	&	\nodata 		&	$1.16\pm0.03$\\
345$-I_{\rm B9}$\tablenotemark{${(f)}$}		&	$1.30\pm0.02$	&	\nodata 		&	$0.80\pm0.02$  	&	\nodata		&	$1.48\pm0.04$ 	&	\nodata\\		

\tablenotetext{$a$}{Ratio between the eastern peak of the radial slice crossing the maximum emission on the eastern lobe, and the western peak of the radial slice crossing the maximum emission on the western lobe (see {\it black/white} and {\it red} profiles in Figure~\ref{fig:Asym}).}
\tablenotetext{$b$}{Ratio between the total flux densities integrated over the eastern and western halves of the image. The image is split at the RA associated with the SN site \citep{rey95}. }
\tablenotetext{$c$}{Ratio between the eastern and western peaks of the radial slice at PA $90^{\circ}$ (see {\it dark green} profile in Figure~\ref{fig:Asym}).}
\tablenotetext{$d$}{Images resolved with a $0\farcs7$ circular beam (see Figure~\ref{fig:Asym}).}
\tablenotetext{$e$}{Diffraction-limited (DL) images (see Figure~\ref{fig:df}).}
\tablenotetext{$f$}{Images derived after subtraction of the model flux density at 672 GHz ($I_{\rm B9}$), scaled to fit the central emission (see Figure~\ref{fig:Asym}).\\ \\ \\}
\end{deluxetable*}
\end{center}
\vspace{0.0mm}

\vspace{-4.0mm}
The effects of the asymmetric shock propagation are likely to emerge in the transition from radio to FIR  rather than at lower frequencies, due to the shorter synchrotron lifetime at higher frequencies. 
To estimate the synchrotron lifetime in the FIR range, we use the approximation that, in a magnetic field of strength $B$, all the radiation of an electron of energy $E$ is emitted only at the critical frequency $\nu_{c}$ \citep{ryb79}.  
Considering the electronÕs orbit is inclined at a pitch angle $\theta$ to the magnetic field, the synchrotron lifetime, $\tau_{e}$, can be derived as a function of $\nu_{c}$ (e.g. \citealp{con92})
\begin{equation}
\tau_{e}\equiv {E \over |dE/dt| }\sim 1.06\times10^{9} (B\,{\rm sin}\theta)^{-3/2} \nu_{c}^{-1/2},
\label{eq:Asym4} 
\end{equation}
where
$\tau_e$ is expressed in years, $B$ in $\mu$G and $\nu$ in GHz. 
As the electrons are expected to have an isotropic distribution of pitch angles, $\langle {\rm sin}^{2}\theta\rangle=2/3$;
for $200\lesssim\nu\lesssim400$ GHz, and assuming a magnetic field strength at the shock front of  $10\lesssim B \lesssim 20$ mG (\citealp{ber06,ber11}), in the radio/FIR transition we can estimate $20\lesssim\tau_{e}\lesssim80$ yr. 
However, since it is likely that sub-diffusive particle transport \citep{kir96} is taking place in regions of the SNR, in conjunction with efficient CR acceleration \citep{glu13}, local 
magnetic-field amplifications could exceed the above limits by at least an order of magnitude
\citep{bel01}. In this scenario, the local $\tau_{e}$ would be of the order of months. 
Therefore,  with regions in the remnant where
electrons might be unable to cross the emission sites within their radiative life-time, the synchrotron emission at FIR frequencies would require the presence of relatively fresh injected and/or re-accelerated electrons to match the emission distribution at lower frequencies.\\

%
%
%
%

\section{Spectral Energy Distribution}
\label{SED}

\begin{figure*}[htp]
\begin{center}
\vspace{0mm}
\advance\leftskip-5mm														  	  %
\includegraphics[trim=6.5mm 5.0mm 10.0mm 10mm, clip=true,width=178mm, angle=0]{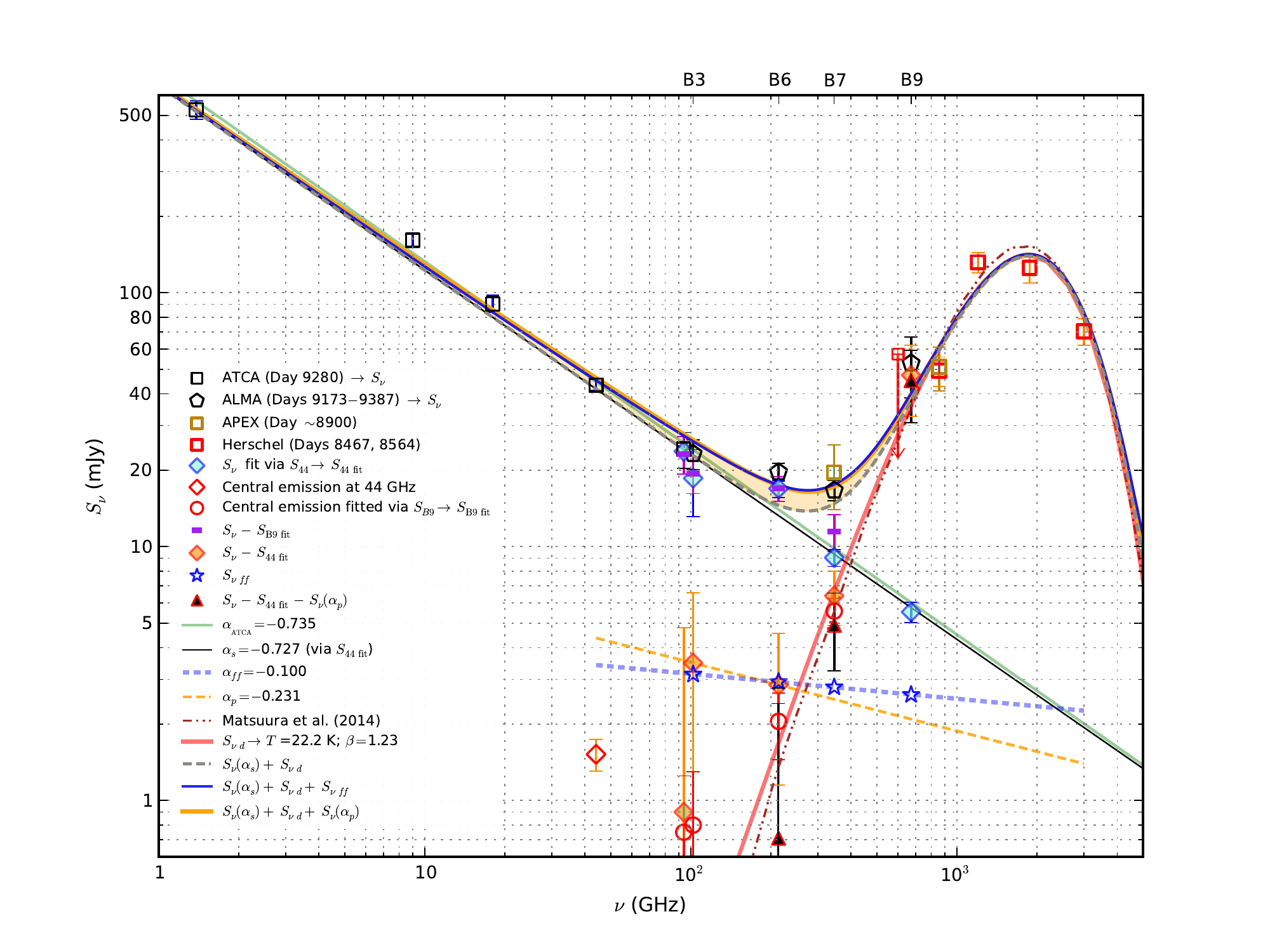}    %
\caption{
Spectral energy distribution (SED) of SNR 1987A from radio to FIR, with data from: ATCA at 1.4 GHz (Zanardo et al. in prep), 9 GHz \citep{ng13}, 18 and 44 GHz \citep{zan13},  and 94 GHz \citep{lak12}; ALMA at 102 GHz (Band 3, B3), 213 GHz (Band 6,  B6), 345 GHz (Band 7, B7) and 672 GHz (Band 9, B9) (this paper);
 the Atacama Pathfinder EXperiment (APEX) at 345 and  857 GHz \citep{lak12b}; the Herschel Space Observatory at 600 -- 3000 GHz \citep{mat11}. 
The brown dash-dot-dotted curve is the  amorphous carbon dust fit for ALMA data and {\it Herschel} observations carried out in 2012 (\citealp{mat14}, submitted).
To match the average epoch of ALMA observations, ATCA data are scaled to day 9280, via  exponential fitting parameters derived for ATCA flux densities from day 8000, as measured at 1.4, 8.6 and 9 GHz (Zanardo et al., in preparation; \citealp{sta14}). 
The hollow red diamond indicates the central emission measured at 44 GHz as reported by \citet{zan13},  scaled to day 9280.
 The difference between possible spectral fits is highlighted in light orange.
See \S~\ref{SED} for a detailed description of this figure.  \\}
\label{fig:SED}
\end{center}
\end{figure*}

The spectral energy distribution for ATCA data from 1.4 to 94 GHz and ALMA data  is shown in Figure~\ref{fig:SED}. To match the average epoch of the ALMA data, the ATCA data are scaled to day 9280, via  exponential fitting parameters derived for ATCA flux densities from day 8000, as measured at 1.4, 8.6 and 9 GHz (Zanardo et al., in preparation; \citealp{sta14}).
Across the transition from radio to FIR frequencies, the observed spectrum consists of the sum of thermal and non-thermal components.

As described in  \S~\ref{Obs}, to identify the dust component of the emission from the inner regions of the remnant, the B9 model flux density, $S_{\rm B9}$, has been scaled to fit the emission measured in the SNR central region at 94--345 GHz ($S_{\rm B9_{\rm \,fit}}$, hollow red circles in Figure~\ref{fig:SED}). The  subtraction of $S_{\rm B9_{\rm \,fit}}$ from the visibilities at 94--345 GHz yields the residual flux densities indicated as $S_{\nu}-S_{\rm B9_{\rm \,fit}}$ (purple bars in Figure~\ref{fig:SED}), as for the images shown in the central column of  Figure~\ref{fig:df}. 
The flux densities $S_{\rm B9_{\rm \,fit}}$ derived in B6 and B7, together with the total integrated flux density measured in B9 ($S_{\rm B9}$), although obtained via a different reduction technique, have been associated by \citet{ind14} 
with dust grains, 
in conjunction with data from {\it Herschel} \citep{mat11} and the Atacama Pathfinder Experiment (APEX; \citealp{lak12b}) (see Figure~\ref{fig:SED}).

Similarly (see \S~\ref{Obs}), to separate the non-thermal emission from that thermal, the 44 GHz model flux density, $S_{44}$, has been scaled to fit the toroidal component of the emission at 94--672 GHz 
(blue/cyan diamonds in Figure~\ref{fig:SED}). 
By fitting  the resulting  $S_{44_{\rm \,fit}}$ components and the ATCA flux densities at 1.4--44 GHz, we obtain, at day 9280, the synchrotron spectral index $\alpha_{s}=-0.727\pm0.020$, with  $S_{\nu}(\alpha_{s})\propto\nu^{\alpha_{s}}$.  
The spectral index measured from 1.4 to 94 GHz, i.e. for ATCA data only, is $\alpha_{_{\rm ATCA}}=-0.735\pm0.028$. While  $\alpha_{s}$ is slightly flatter than $\alpha_{_{\rm ATCA}}$, both values are consistent with the progressive flattening of the radio spectrum measured since day $\sim5000$ \citep{zan10}.
The subtraction of $S_{44_{\rm \,fit}}$ from the visibilities at 94--672 GHz, gives the residual flux densities $S_{\nu}-S_{44_{\rm \,fit}}$ (red/orange diamonds in Figure~\ref{fig:SED}), as for the images shown in the right column of  Figures~\ref{fig:df}. 
In B3, B6 and B7,  $S_{\nu}-S_{44_{\rm \,fit}} > S_{\rm B9_{\rm \,fit}}$, i.e. the residuals exceed the emission expected from the  dust. 

While the subtraction of the flux densities is inevitably affected by errors (see Table~\ref{tab02}), 
given that the $I_{44}-$subtracted images have a S/N$\,\gtrsim7\, \sigma$  and the residual emission appears primarily located westwards of the optical ejecta (see \S~\ref{Subtract}),
we investigate the nature of this emission excess as:  
(1) free-free emission from an ionized fraction of the inner ejecta; 
(2) synchrotron emission from a compact source located in the inner regions of the remnant;
(3) emission from grains of very cold dust.

%
%

\subsection{Free-Free Emission}
\label{ff}

To estimate the free-free radiation in the SNR as imaged with ALMA, 
we hypothesize an ionized portion of the ejecta as an approximately spherical region, located inside the ER. 
Using the beam size of the super-resolved $I_{44}-$subtracted image at 102 GHz as an upper limit, we consider the radius of the spherical region up to  
$R_{s}\sim0\farcs40$ ($\approx3.08\times10^{17}$ cm).
Such radius covers the extent of the inner ejecta as imaged in the optical \citep{lar13}, and stretches to the likely radius of the reverse shock, qualitatively identified with the inner edge of the emission over the ER (see Figure~\ref{fig:B7_Hbl}). 
Given that the pre-supernova mass has been estimated between $14\,M_{\sun}$ and $20\,M_{\sun}$ (\citealp{sma09}),
if we assume that the H{\sc ii} region in the ejecta has a uniform density $\rho_{ej}(r)= 3\, M_{\rm HII} (4 \pi r^{3})^{-1}$, 
we set  $0.7\,M_{\sun} \lesssim M_{\rm HII}  \lesssim 2.5\,M_{\sun}$. 
The lower limit, $M_{\rm HII}\sim0.7\,M_{\sun}$, i.e. $\rho_{ej}(r=R_{s})=1.1\times10^{-20}$,
represents a partial ionization of the 
ejecta by X-ray flux, either within the inner region or on the outer layer. 
The upper limit, $M_{\rm HII} \sim 2.5\,M_{\sun}$,  i.e. 
$\rho_{ej}(r=R_{s})=4.1\times10^{-20}$ g cm$^{-3}$ 
matches the density model by \citet{bli00} scaled to the current epoch 
(see Figure 21 in \citealp{fra13}), and corresponds to complete ionization 
of the H and He mass within $R_{s}$. 

The optical depth associated with the H{\sc ii} region along the line of sight ({\it los}) can be estimated as
\begin{equation}
\tau_{ff}\approx3.28\times10^{-7} \,T_{4}^{-1.35} \nu^{-2.1} N_{e}^{2} R_{s},
\label{eq:SED1}  
\end{equation}
where $T_{4}=T_{e}/{\rm (10^{4} \,K)}$, $\nu$ is in GHz, $N_{e}$ is in cm$^{-3}$ and $R_{s}\approx \int_{\rm los} \mathrm{d} l$ is in pc. 
For $T_{4}\sim 1$, given the emission measure 									%
$4.6\times10^{6}\lesssim {\rm EM} \lesssim5.9\times10^{7}$ cm$^{-6}$ pc,				%
where ${\rm EM}=N_{e}^2 R_{s}$, at frequencies $102\lesssim\nu\lesssim672$ GHz		%
 the emission becomes nearly transparent as										%
$2.28\times10^{-6}\lesssim\tau_{ff}\lesssim1.2\times10^{-3}$. 							%
The flux associated with the ionized component of the ejecta, can then be derived as 
\begin{equation}
S _{\nu_{ff}} \propto  \tau_{ff}(\nu)   {2 k T  \over \lambda^2} \Omega \propto \nu^{-0.1}.	
\label{eq:SED2}  
\end{equation}

Considering the solid angle $\Omega$ subtended by the same radius, $R_{s}$, at all frequencies,
for  the lower limit $M_{\rm HII}\approx0.7\,M_{\sun}$,  i.e. $N_{e}=6.8\times10^{3}$ cm$^{-3}$, 
$S _{\nu_{ff}} \sim (S_{\nu}-S_{44_{\rm \,fit}})$ at 102 and 213 GHz, 
as Eq.~\ref{eq:SED2} yields 
$S_{102_{ff}}\sim3.9$ mJy, $S_{213_{ff}}\sim3.2$ mJy, $S_{345_{ff}}\sim3.1$ mJy and $S_{672_{ff}}\sim2.9$ mJy. 
These  $S _{\nu_{ff}}$ values  (blue stars in in Figure~\ref{fig:SED}) would well fit the SED  (see blue fit  in Figure~\ref{fig:SED}). 
If the H{\sc ii} density is considerably higher, as given by the upper limit $M_{\rm HII}\approx2.5\,M_{\sun}$, the derived $S _{\nu_{ff}}$ fluxes would exceed the emission residuals by an order of magnitude.


As a constraint to the free-free emission component,
the hypothesized H{\sc ii} region would also produce optical and near-IR H{\sc i}
recombination lines. The resultant H${\alpha}$ flux  can be estimated as 
\begin{equation}
S_{\rm H\alpha} \approx \epsilon  \, N_{e} N_{+} \, \biggl({{4}\over{3}} \pi R_{s}^3 \biggr)  {{f} \over {4 \pi d^{2}}},
\label{eq:RRL1}
\end{equation}
where 
$\epsilon  \propto \alpha_{A}$ is the emissivity of the H{\sc $\alpha$} line per unit volume, 
with $\alpha_{A}=\alpha_{A}(T)={\sum_{n} \alpha_{n}(T)}$
 the total recombination coefficient,  
$f$ is the volume filling factor and $d$ is the distance to the source.
If one adopts 
$\epsilon\approx 3.53\times10^{-25}\,T_{4}^{-0.92}$ erg cm$^{-3}$ s$^{-1}$ for $T_{4}\simeq0.5$ (e.g. \citealp{sto95}),  
the number of ions $N_{+}=N_{e} = 6.8\times10^{3}$ cm$^{-3}$  
as for the lower  limit assumed for the density of the  H{\sc ii} region,  and  $R_{s}\approx3.08\times10^{17}$ cm, 
for $f\sim1$ one obtains $S_{\rm H\alpha}\approx 1.26\times10^{-11}$ erg cm$^{2}$ s$^{-1}$. 
To match the H$\alpha$ flux from the core 
measured by \citet{fra13} (see Figure 8 therein) at $\sim1.4\times10^{-14}$ erg cm$^{2}$ s$^{-1}$ on day 9000, 
the estimated $S_{\rm H\alpha}$  
would have to undergo $7.4$ magnitudes of extinction.
Such extinction is  possible but  improbable. 
We note that the above $S_{{\rm H}\alpha}$ also exceeds the H$\alpha$ emission by the reverse shock,  measured at $\sim2.0\times10^{-13}$ erg cm$^{2}$ s$^{-1}$ on day 9000, and associated with a density 
 $250\lesssim N_{+}\lesssim750$ cm$^{-3}$ \citep{fra13}. 

As regards the magnitude of the flux from the Br-$\gamma$ line,
the model estimate by  \citet{kja10} on day 6840  is a factor of $\sim10^{-2}$ smaller than the $S_{\rm H\alpha}$ derived from Eq.~\ref{eq:RRL1},
while this is mainly associated with emission from the hot spots in the ER.

In the absence of external X-ray heating,
both Êthe heating and Êionization would be powered by radioactive decays in the core.
The resultant flux, estimated via models as in \citet{koz98},  would be several orders of magnitude smaller than the  lower limit for $S_{\nu_{ff}}$.  

Another possible source of ionizing emission is a pulsar wind nebula (PWN)
 in the SNR interior. The properties of any PWN are very uncertain (see \S~\ref{PWN--NS}), but there is the 
 expectation that line emission would accompany free-free emission also in this case.


\subsection{Flat-Spectrum Synchrotron Emission}
\label{sync2}
 
A flat spectral index could also be attributed to a second synchrotron component. 
As shown in  Figure~\ref{fig:SED}, within $102\le\nu\le672$ GHz, a synchrotron component with spectral index $\alpha_{p}=-0.231$ fits the residuals ($S_{\nu}-S_{44_{\rm \,fit}}$) at 102 and 213 GHz. 
Synchrotron emission with  $-0.4\lesssim\alpha\lesssim -0.1$ fits the spectrum in the radio/FIR transition and could originate from a compact source near the center of the SNR. In the case of a central pulsar, the synchrotron emission would be generated by the shocked magnetized particle wind \citep{gae06}. 
The scenario of a synchrotron-emitting PWN in the inner SNR is explored in \S~\ref{PWN--NS}.


\subsection{Dust Emission}
\label{dust}

If the excess emission is due to an additional synchrotron component, $S_{\nu}(\alpha_{p})\propto\nu^{-0.2}$, this would
provide a constraint to the net dust emission, $S_{\nu_{d}}$, in ALMA data. 
The subtraction $S_{\nu}-S_{44_{\rm \,fit}}-S_{\nu}(\alpha_{p})$, where 
$S_{\nu_{p}}=S_{0_{p}}(\nu/ \nu_{0})^{\alpha_{p}}$
with $S_{0_{p}}=3.5\pm3.1$ mJy as from Table~\ref{tab02},
leads to
$S_{213_{d}}\sim0.7$ mJy,		
$S_{345_{d}}\sim4.9$ mJy and 		
$S_{672_{d}}\sim45.0$ mJy (black triangles in Figure~\ref{fig:SED}).		
We note that $S_{672_{d}}$ coincides with the integrated flux density of the central feature of the related image, which extends over the inner ejecta as seen with {\it HST} (see Figure~\ref{fig:B7_Hbl}). 
The net dust can be fitted via a modified Planck curve of thermal emission, as 
\begin{equation}
S_{\nu_{d}}(\beta,T)=  {M_{\rm dust} \over d^{2}} \,  \kappa_{\nu} \, B_{\nu}(T)\propto  \nu^{\,\beta}\,B_{\nu}(T),
\label{eq:SED4}	
\end{equation}
where
$M_{\rm dust}$ is the dust mass,
 $\kappa_{\nu} = 3 \,Q_{\nu}  / 4 \rho \,a_{\nu}$
is the dust mass absorption coefficient, with $Q_{\nu}$ the absorption efficiency  for spherical grains of density $\rho$ and radius $a_{\nu}$,
$1\lesssim\beta\lesssim2$ for insterstellar dust (e.g. \citealp{cor14} and references therein) and
$B_{\nu}(T)$ is the Planck function.
As shown in Figure~\ref{fig:SED}, the best fit  of both ALMA fluxes, as reported in this paper, and  {\it Herschel}  fluxes, as from \citet{mat11},  yields  $\beta=1.23$ 
and dust temperature $T=22.2$ K. 
The thermal peak of the SED  has been previously fitted (\citealp{mat11,lak11, lak12,lak12b}) with temperatures estimated between 17 and 26 K, while \citet{ind14} fit amorphous carbon dust at $T=26\pm3$ K.

The sum of the main synchrotron component, $S_{\nu}\propto\nu^{-0.73}$, and the emission component from dust grains at $T\sim 22$ K is lower than the measured ALMA flux densities at 213 and 345 GHz (see dashed gray fit in Figure~\ref{fig:SED}). 
To match the emission excess of $\sim3$ mJy in this frequency range, we could also 
postulate a second dust component.
This would require very cold dust at $T \lesssim5$ K, 
i.e. at temperatures where the assumption of either amorphous carbon or silicates leads to dust 
masses implausibly large for physically realistic grains. 
In particular, as from Eq. \ref{eq:SED4}, at 345 GHz  a flux density of $\sim3$ mJy would require dust at $T\sim3$ K with $M_{\rm dust}\sim50 \,M_{\sun}$, as obtained by using $k_{\nu}\sim2.5$ cm$^{2}$ g$^{-1}$ for amorphous carbon (\citealp{zub96,zub04}).
We note that warmer dust at $T\sim$ 180 K  has been  identified by  \citet{dwe10} in the ER,  where the dust grains are likely collisionally heated by the expanding radiative shocks \citep{bou06}.\\

\section{Spectral Index Variations}
\label{SI}

%
%
\begin{figure*}[htp]
\begin{center}
\vspace{3mm}
\advance\leftskip-10mm
\includegraphics[trim=5.mm 80.0mm 5.0mm 0.0mm, clip=true,width=194mm, angle=0]{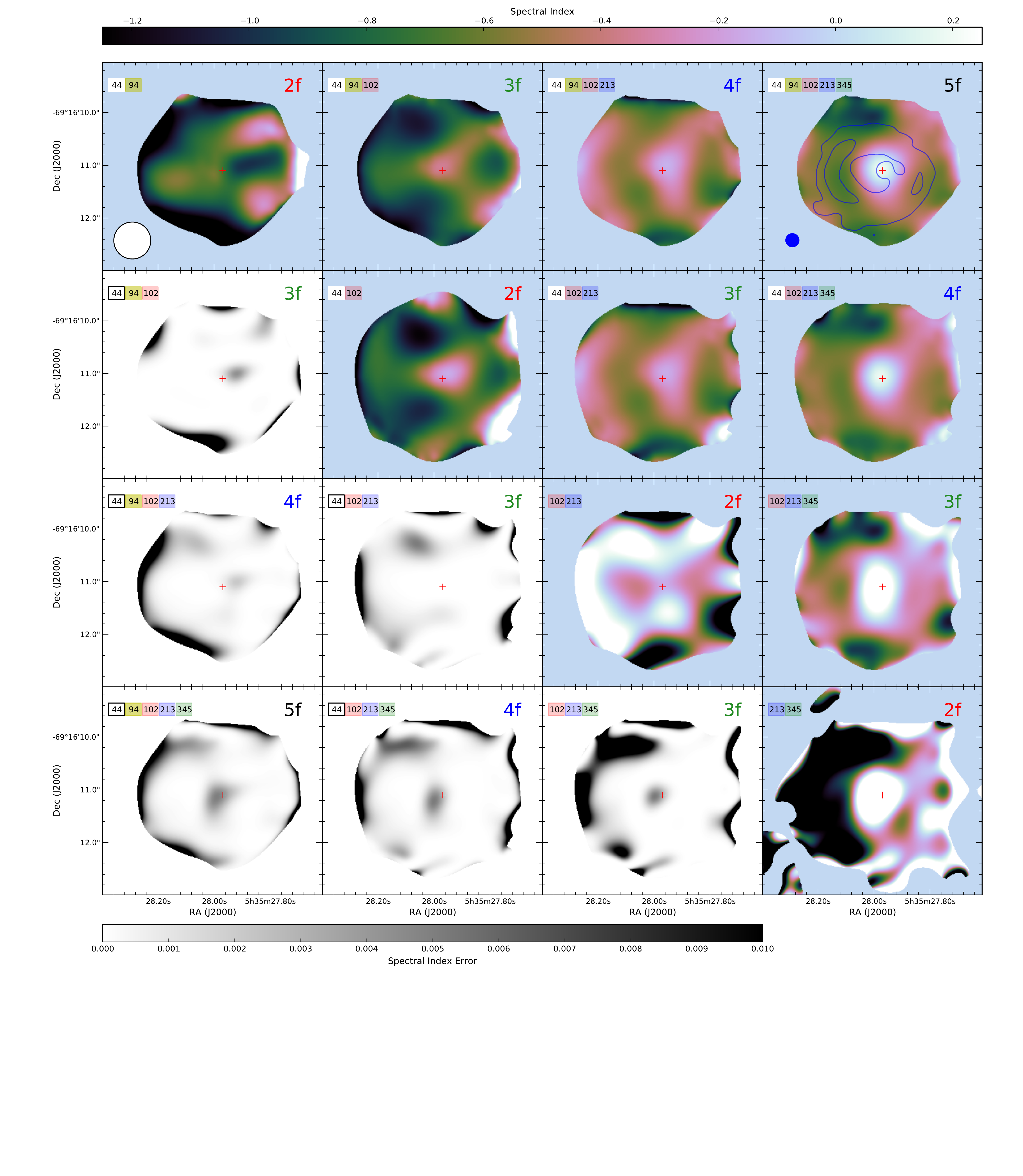}
\caption{
In color scale are maps of the spectral index distribution in SNR 1987A, as derived from images  at 44, 94, 102, 213, and 345 GHz. The source images are reduced with identical procedure and restored with a circular beam of $0\farcs8$. 
To match the average epoch of ALMA observations, ATCA data at 44 and 94 GHz are scaled to day 9280, via  exponential fitting parameters derived for ATCA flux densities from day 8000, as measured at  8.6 and 9 GHz (Zanardo et al., in preparation). 
The spectral maps are derived from flux densities at 2, 3, 4, and 5 frequencies, and labelled as {\it2f}, {\it3f}, {\it4f}, and {\it5f}, respectively (top right corner).  The frequencies used in each map are indicated, in GHz, on the top left corner.   Image regions below the highest rms noise level are masked. The upper limit of the spectral index color scale is set to $\alpha=0.25$, therefore map regions with spectral indices greater than 0.25 appear in white. 
In grey scale are maps of the error associated with $\alpha$ in the power-law fit, $S_{\nu}\sim\nu^{\alpha}$, used for spectral maps of flux densities at 3, 4, and 5 different frequencies, and labelled as {\it3f}, {\it4f}, and {\it5f}, respectively. To emphasize the error distribution, while using a linear gradient of the grey color scale, map regions with errors greater that 0.01 appear in black. In the top right map, the contours of the 44 GHz image, resolved with a $0\farcs25$  circular beam, are shown at the $15\%$ and $60\%$ emission levels (in blue).
The red cross indicates the VLBI position of SN 1987A as determined by \citet{rey95} [RA $05^{\rm h}\;35^{\rm m}\;27\fs968$, Dec $-69^{\circ}\;16'\;11\farcs09$ (J2000)].}
\label{fig:spectral_map}
\end{center}
\end{figure*} 
%
%

%
%
\begin{figure*}[htp]
\begin{center}
\vspace{2mm}
\advance\leftskip-3mm								%
\includegraphics[width=178.0mm, angle=0]{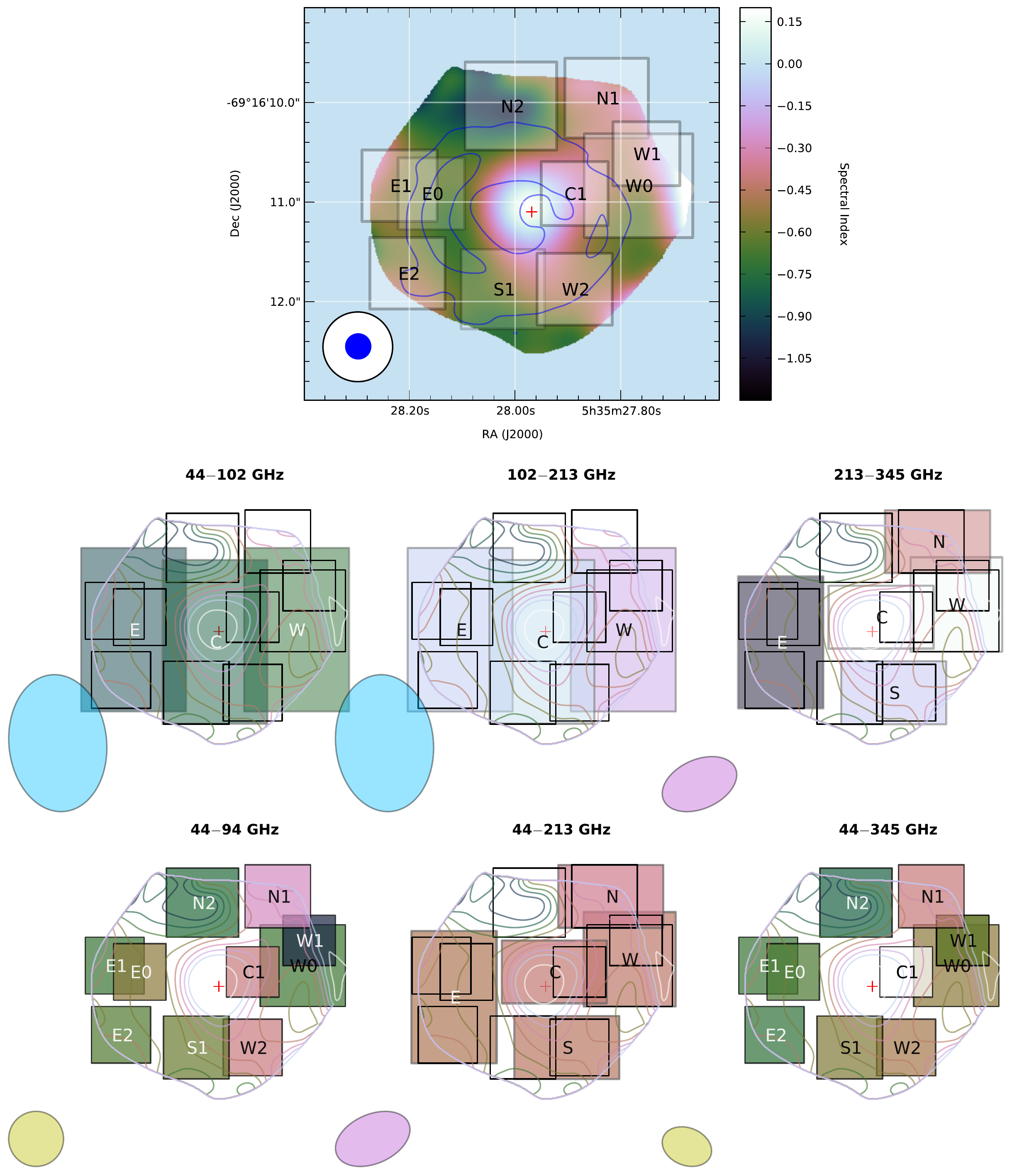}  		%
\caption{ 
{\it Top} $-$
Four-frequency (44--94--213--345 GHz) spectral index intensity map with an angular resolution of $0\farcs8$, where ATCA data at 44 and 94 GHz are scaled to day 9280. 
The map is superimposed with the contours of the 44 GHz image, resolved with a $0\farcs25$  circular beam, at the $15\%$ and $60\%$ emission levels (in blue). 
The regions used to derive the spectral indices $\alpha_{TT}$, via {\it T-T}  plots, are indicated by grey/white squares.
{\it Bottom} $-$
The {\it T-T}  plots are applied to 6 pairs of frequencies:  $44-102$ GHz, $102-213$ GHz,  $213-345$ GHz,  $44-94$ GHz, $44-213$ GHz, and  $44-345$ GHz. 
The controlling beam associated with each pair of frequencies is shown on the lower left corner of the images. 
The {\it T-T}  boxes are color-coded according to the resultant $\alpha_{TT}$  value.   
For image pairs of angular resolution lower than $0\farcs8$, larger {\it T-T}  boxes are used as outlined.		%
All {\it T-T}   boxes are drawn on top of the 4-frequency spectral map, highlighted by contours spaced at $\Delta\alpha=0.1$, which are colored according to the associated spectral index.
The red cross indicates the VLBI position of SN 1987A as determined by \citet{rey95}.
The $\alpha_{TT}$  values resulting from the six frequency pairs are detailed  
in Table~\ref{tab_alpha}.}
\label{fig:TT_boxes_color}
\end{center}
\end{figure*} 
%
%

%
%
\begin{center}
\begin{deluxetable*}{ccccccccc}
\tablecaption{Regional Spectral Indices \label{tab_alpha}} 
\tablewidth{0mm}
\tablecolumns{9}
\tablehead{
\colhead{} 													&  
\colhead{} 													&
\multicolumn{6}{c}{$\alpha_{TT} (S_{1},S_{2})$\tablenotemark{$^{(a)}$}\tablenotemark{$^{(e)}$}} 	& 
\colhead{$\alpha_{M_\mu}(S_{1,2,3,4})$\tablenotemark{$^{(b)}$}\tablenotemark{$^{(e)}$}}   		\\  
\cmidrule(lr){3-8} 
\cmidrule(lr){9-9} 							
\colhead{Region\tablenotemark{$^{(c)}$}} 							& 
\colhead{Fit\tablenotemark{$^{(d)}$}} 								&
\colhead{$_{(44,102)}$} 										& 
\colhead{$_{(102,213)}$}										& 
\colhead{$_{(213,345)}$}	 									& 
\colhead{$_{(44,94)}$} 										& 
\colhead{$_{(44,213)}$} 										& 
\colhead{$_{(44,345)}$} 										& 
\colhead{$_{(44,94,213,345)}$}
}				
\startdata	
N			&	$_{y}$		&	\nodata		&	\nodata 			&	$-0.33\pm0.02$ 	&	\nodata			&	$-0.38\pm0.03$ 		&	\nodata 	          	   &	\nodata\\
			&	$_{ODR}$		&	\nodata		&	\nodata 			&	$-0.41\pm0.02$ 	&	\nodata			&	\nodata 			&	\nodata 			   &	\nodata\\
			&	$_{Abs}$		&	\nodata		&	\nodata			&	$-0.40\pm0.02$ 	&	\nodata			&	$-0.37\pm0.03$ 		&	\nodata 			   &	\nodata\\
			&	$_{(0,0)}$		&	\nodata		&	\nodata 			&	$-0.05\pm0.04$ 	&	\nodata			&	$-0.43\pm0.05$ 		&	\nodata 			   &	\nodata\\
N1			&	$_{y}$		&	\nodata		&	\nodata 			&	\nodata 		&	$-0.28\pm0.08$		&	\nodata 	  		&	$-0.39\pm0.02$ 	   	   &	$-0.39\pm0.09$\\
			&	$_{ODR}$		&	\nodata		&	\nodata 			&	\nodata 	   	&	$-0.31\pm0.09$		&	\nodata 	  		&	$-0.42\pm0.02$ 	   	   &	\nodata\\
			&	$_{Abs}$		&	\nodata		&	\nodata 			&	\nodata 	   	&	$-0.27\pm0.08$		&	\nodata 	  		&	$-0.35\pm0.03$ 	   	   &	\nodata\\
			&	$_{(0,0)}$		&	\nodata		&	\nodata 			&	\nodata 	   	&	$-0.36\pm0.14$		&	\nodata 	  		&	$-0.40\pm0.04$ 	   	   &	\nodata\\
N2			&	$_{y}$		&	\nodata		&	\nodata			&	\nodata 	   	&	$-0.73\pm0.03$		&	\nodata 			&	$-0.81\pm0.08$ 	   	   &	$-0.85\pm0.14$\\
			&	$_{ODR}$		&	\nodata		&	\nodata			&	\nodata 	   	&	$-0.76\pm0.03$		&	\nodata 	   		&	\nodata 	   		   &	\nodata\\
			&	$_{Abs}$		&	\nodata		&	\nodata			&	\nodata 	   	&	$-0.77\pm0.03$		&	\nodata 	   		&	\nodata 	   		   &	\nodata\\
			&	$_{(0,0)}$		&	\nodata		&	\nodata			&	\nodata 	   	&	$-0.84\pm0.04$		&	\nodata 	   		&	$-0.89\pm0.07$ 	   	   &	\nodata\\
\cmidrule(lr){1-9}		
E			&	$_{y}$		&	$-0.95\pm0.02$	&	$-0.07\pm0.02$ 		&	$-1.05\pm0.04$ 	&	\nodata			&	$-0.48\pm0.02$ 		&	\nodata 	   	  	   &	\nodata\\
			&	$_{ODR}$		&	$-0.95\pm0.01$	&	$-0.07\pm0.03$ 		&	\nodata 		&	\nodata			&	$-0.60\pm0.05$ 		&	\nodata 	   	  	   &	\nodata\\
			&	$_{Abs}$		&	$-0.95\pm0.02$	&	$-0.06\pm0.03$ 		&	$-1.12\pm0.04$ 	&	\nodata			&	$-0.47\pm0.02$ 		&	\nodata 	   	  	   &	\nodata\\
			&	$_{(0,0)}$		&	$-0.97\pm0.02$	&	$-0.03\pm0.03$ 		&	$0.11\pm0.07$ 	&	\nodata			&	$-0.46\pm0.02$ 		&	\nodata 	   		   &	\nodata\\
E0			&	$_{y}$		&	\nodata		&	\nodata			&	\nodata 	   	&	$-0.55\pm0.03$		&	\nodata 	   		&	$-0.68\pm0.03$ 	  	   &	$-0.60\pm0.03$\\	
			&	$_{ODR}$		&	\nodata		&	\nodata			&	\nodata 	   	&	$-0.59\pm0.03$		&	\nodata 	   		&	$-0.72\pm0.04$ 	            &	\nodata\\
			&	$_{Abs}$		&	\nodata		&	\nodata			&	\nodata 	   	&	$-0.54\pm0.03$		&	\nodata 	   		&	$-0.66\pm0.03$ 	   	   &    \nodata\\
			&	$_{(0,0)}$		&	\nodata		&	\nodata			&	\nodata 	   	&	$-0.69\pm0.07$		&	\nodata 	   		&	$-0.71\pm0.06$      	   &	\nodata\\
E1			&	$_{y}$		&	\nodata		&	\nodata			&	\nodata 	   	&	$-0.65\pm0.04$		&	\nodata 	   		&	$-0.66\pm0.04$  	      	   &	$-0.52\pm0.13$\\	
			&	$_{ODR}$		&	\nodata		&	\nodata			&	\nodata 	   	&	$-0.68\pm0.04$		&	\nodata 	   		&	$-0.68\pm0.04$ 	   	   &	\nodata\\
			&	$_{Abs}$		&	\nodata		&	\nodata			&	\nodata 	   	&	$-0.70\pm0.05$		&	\nodata 	   		&	$-0.65\pm0.04$ 	   	   &	\nodata\\
			&	$_{(0,0)}$		&	\nodata		&	\nodata			&	\nodata 	   	&	$-0.78\pm0.08$		&	\nodata 	   		&	$-0.68\pm0.04$ 	   	   &	\nodata\\
E2			&	$_{y}$		&	\nodata		&	\nodata			&	\nodata 	   	&	$-0.60\pm0.03$		&	\nodata 	   		&	$-0.73\pm0.03$ 	   	   &	$-0.55\pm0.12$\\
			&	$_{ODR}$		&	\nodata		&	\nodata			&	\nodata 	   	&	$-0.63\pm0.03$		&	\nodata 	   		&	$-0.74\pm0.02$ 	   	   &	\nodata\\
			&	$_{Abs}$		&	\nodata		&	\nodata			&	\nodata 	   	&	$-0.65\pm0.03$		&	\nodata 	   		&	$-0.72\pm0.03$ 	   	   &	\nodata\\
			&	$_{(0,0)}$		&	\nodata		&	\nodata			&	\nodata 	   	&	$-0.74\pm0.04$		&	\nodata 	   		&	$-0.73\pm0.03$ 	   	   &	\nodata\\
\cmidrule(lr){1-9}			
S			&	$_{y}$		&	\nodata		&	\nodata 			&	$-0.05\pm0.05$ 	&	\nodata			&	$-0.45\pm0.02$ 		&	\nodata 	   		    &	\nodata\\	
			&	$_{ODR}$		&	\nodata		&	\nodata 			&	$-0.10\pm0.04$ 	&	\nodata			&	\nodata 			&	\nodata 	    	 	    &	\nodata\\	
			&	$_{Abs}$		&	\nodata		&	\nodata 			&	$-0.23\pm0.03$ 	&	\nodata			&	$-0.44\pm0.02$ 		&	\nodata 	   		    &	\nodata\\	
			&	$_{(0,0)}$		&	\nodata		&	\nodata 			&	$-0.22\pm0.03$ 	&	\nodata			&	$-0.53\pm0.07$		&	\nodata 	   		    &	\nodata\\	
S1			&	$_{y}$		&	\nodata		&	\nodata			&	\nodata 	   	&	$-0.61\pm0.02$		&	\nodata 	  		&	$-0.52\pm0.07$ 	    	    &	$-0.63\pm0.07$\\
			&	$_{ODR}$		&	\nodata		&	\nodata			&	\nodata 	   	&	\nodata			&	\nodata 	  		&	\nodata 	   	    	&	\nodata\\
			&	$_{Abs}$		&	\nodata		&	\nodata			&	\nodata 	   	&	$-0.60\pm0.02$		&	\nodata 	  		&	$-0.56\pm0.07$ 	    	    &	\nodata\\
			&	$_{(0,0)}$		&	\nodata		&	\nodata			&	\nodata 	   	&	$-0.95\pm0.18$		&	\nodata 	  		&	$-0.67\pm0.09$ 	             &	\nodata\\
\cmidrule(lr){1-9}			
W			&	$_{y}$		&	$-0.77\pm0.02$	&	$-0.11\pm0.03$ 		&	$0.25\pm0.04$ 	&	\nodata			&	$-0.40\pm0.06$ 		&	\nodata 	   	  	   &	\nodata\\	
			&	$_{ODR}$		&	$-0.78\pm0.02$	&	$-0.11\pm0.02$ 		&	$0.17\pm0.02$ 	&	\nodata			&	$-0.43\pm0.06$ 		&	\nodata 	   		   &	\nodata\\	
			&	$_{Abs}$		&	$-0.77\pm0.02$	&	$-0.10\pm0.03$ 		&	$0.08\pm0.04$ 	&	\nodata			&	$-0.43\pm0.06$ 		&	\nodata 	   	 	   &	\nodata\\	
			&	$_{(0,0)}$		&	$-0.76\pm0.02$	&	$-0.18\pm0.03$ 		&	$0.16\pm0.03$ 	&	\nodata			&	$-0.49\pm0.05$ 		&	\nodata 	   	 	   &	\nodata\\	
W0			&	$_{y}$		&	\nodata		&	\nodata			&	\nodata 		&	$-0.69\pm0.03$		&	\nodata 	   		&	$-0.49\pm0.09$ 	  	   &	$-0.53\pm0.05$\\	  
			&	$_{ODR}$		&	\nodata		&	\nodata		 	&	\nodata 	   	&	$-0.77\pm0.04$		&	\nodata 	   		&	$-0.57\pm0.10$ 	    	   &	\nodata\\	
			&	$_{Abs}$		&	\nodata		&	\nodata			&	\nodata 	   	&	\nodata			&	\nodata 	   		&	$-0.54\pm0.10$ 	   	   &	\nodata\\	
			&	$_{(0,0)}$		&	\nodata		&	\nodata			&	\nodata 	   	&	$-0.66\pm0.03$		&	\nodata 	   		&	$-0.49\pm0.11$ 	   	   &	\nodata\\	
W1			&	$_{y}$		&	\nodata		&	\nodata			&	\nodata 	   	&	$-0.93\pm0.09$		&	\nodata 	   		&	$-0.61\pm0.09$ 	   	   &	$-0.37\pm0.11$\\
			&	$_{ODR}$		&	\nodata		&	\nodata			&	\nodata 	   	&	$-0.97\pm0.08$		&	\nodata 	   		&	\nodata 	   	  	   &	\nodata\\		 
			&	$_{Abs}$		&	\nodata		&	\nodata			&	\nodata 	   	&	$-0.99\pm0.08$		&	\nodata 	   		&	$-0.60\pm0.09$ 	   	   &   \nodata\\	 
			&	$_{(0,0)}$		&	\nodata		&	\nodata			&	\nodata 	   	&	$-0.75\pm0.10$		&	\nodata 	   		&	$-0.51\pm0.15$ 	   	   &	\nodata\\	 
W2			&	$_{y}$		&  	\nodata		&	\nodata			&	\nodata 	   	&	$-0.40\pm0.11$		&	\nodata 	  		&	$-0.50\pm0.02$ 	   	   &	$-0.47\pm0.11$\\
			&	$_{ODR}$		&  	\nodata		&	\nodata			&	\nodata 	   	&	$-0.44\pm0.11$		&	\nodata 	  		&	$-0.52\pm0.02$ 	   	   &	\nodata\\	
			&	$_{Abs}$		&  	\nodata		&	\nodata			&	\nodata 	   	&	$-0.35\pm0.12$		&	\nodata 	  		&	$-0.49\pm0.02$ 	            &	\nodata\\	
			&	$_{(0,0)}$		&  	\nodata		&	\nodata			&	\nodata 	   	&	$-0.52\pm0.14$		&	\nodata 	  		&	$-0.51\pm0.02$ 	   	   &	\nodata\\
\cmidrule(lr){1-9}					
C			&	$_{y}$		&	$-0.86\pm0.08$	&	$-0.00\pm0.04$ 		&	$1.20\pm0.04$ 	&	\nodata			&	$-0.43\pm0.02$ 		&	\nodata 	   	   	   &	\nodata\\
			&	$_{ODR}$		&	$-0.90\pm0.08$	&	$-0.00\pm0.03$ 		&	$1.97\pm0.09$ 	&	\nodata			&	\nodata 			&	\nodata 	   	 	   &	\nodata\\
			&	$_{Abs}$		&	$-0.86\pm0.08$	&	$-0.00\pm0.03$ 		&	$2.25\pm0.11$ 	&	\nodata			&	$-0.44\pm0.01$		&	\nodata 	   		   &	\nodata\\
			&	$_{(0,0)}$		&	$-0.84\pm0.08$	&	$-0.09\pm0.04$ 		&	\nodata 		&	\nodata			&	$-0.43\pm0.02$ 		&	\nodata 	   		   &	\nodata\\
C1			&	$_{y}$		&	\nodata		&	\nodata			&	\nodata 	   	&	$-0.40\pm0.08$		&	\nodata 	   		&	\nodata 	   		   &	$-0.01\pm0.06$\\
			&	$_{ODR}$		&	\nodata		&	\nodata			&	\nodata 	 	&	\nodata			&	\nodata 	   		&	\nodata 	   		   &	\nodata\\	
			&	$_{Abs}$		&	\nodata		&	\nodata			&	\nodata   		&	$-0.47\pm0.08$		&	\nodata 	   		&	\nodata 	   		   &	\nodata\\	
			&	$_{(0,0)}$		&	\nodata		&	\nodata			&	\nodata   		&	$-0.73\pm0.15$		&	\nodata 	   		&	\nodata 	   		   &	\nodata\\	
\enddata
\tablenotetext{$a$}{The spectral indices $\alpha_{TT}$  are derived from  5 images at 44, 94, 102, 213, and 345 GHz. The images are analyzed in 6 frequency pairs, as indicated in Figure~\ref{fig:TT_boxes_color}.}
\tablenotetext{$b$}{The median spectral index, $\alpha_{M_\mu}$, is derived by fitting a Gaussian to the histogram of the $\alpha_{M}$ values resulting in the spectral map obtained from images at 44, 94, 213, and 345 GHz, in the corresponding {\it T-T}  region (see top map in Figure~\ref{fig:TT_boxes_color}).  }
\tablenotetext{$c$}{The regions selected for the {\it T-T}  plots are designated in Figure~\ref{fig:TT_boxes_color}.}
\tablenotetext{$d$}{The given spectral indices are derived from different linear interpolations: least-square fit from vertical squared errors, $\alpha_{y}$; orthogonal distance regression, $\alpha_{ODR}$;  robust fitting from the square root of absolute residuals, $\alpha_{Abs}$; linear regression with forced zero interception $(S_{1}=0,S_{2}=0)$, $\alpha_{(0,0)}$.} 
\tablenotetext{$e$}{The errors on $\alpha_{TT}$  and $\alpha_{M_\mu}$ are the $1\,\sigma$ error on the slope of the linear fit  combined with the uncertainty in the flux calibration.}
\end{deluxetable*}
\end{center}
\vspace{0.0mm}
%
%

%
%
\begin{figure}[htp]
\begin{center}
\vspace{0mm}
\advance\leftskip-9mm
\includegraphics[trim=0mm 5.0mm 0.0mm 11.0mm, clip=true,width=80mm, width=105mm, angle=0]{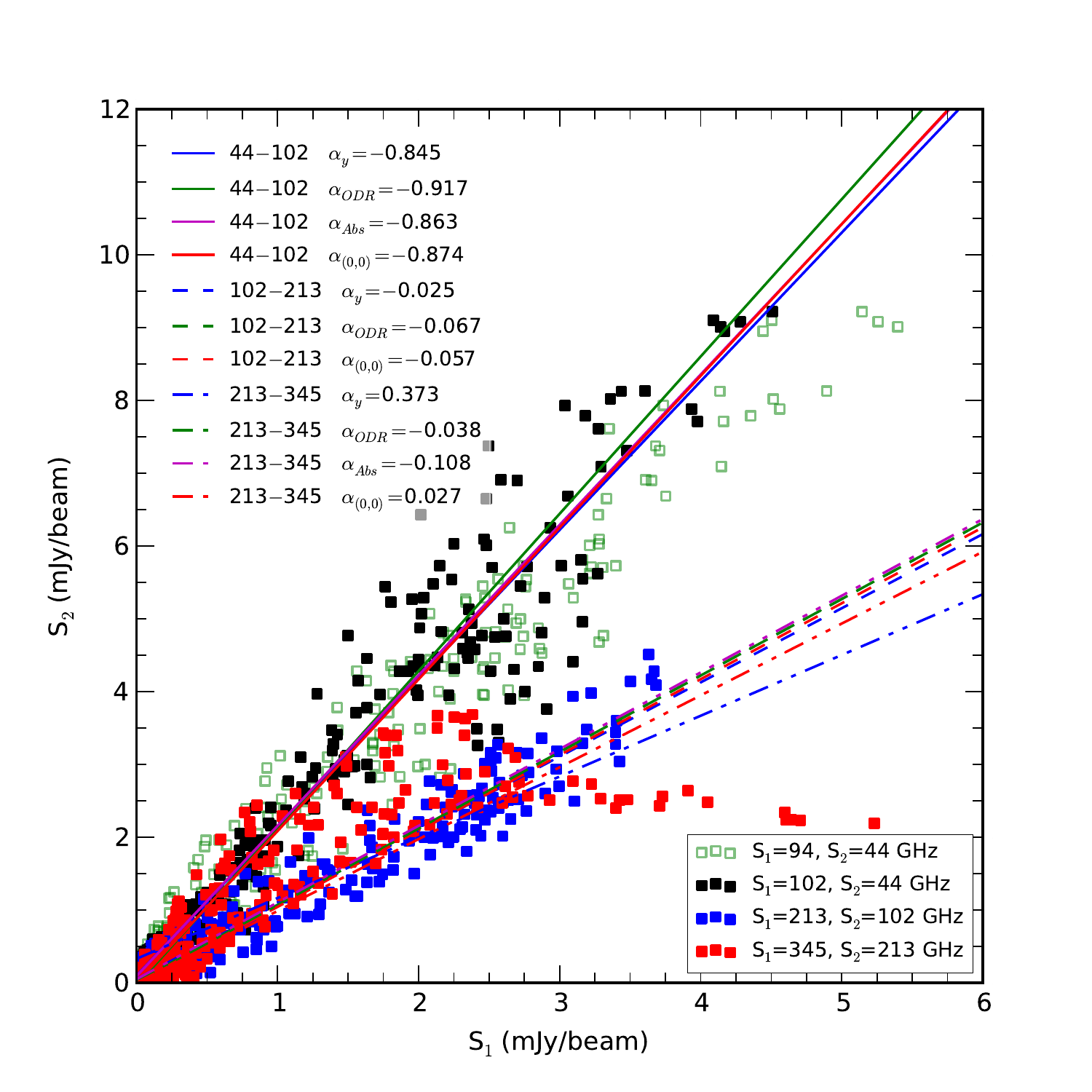}
\caption{
Temperature-temperature ({\it T-T})  plot of the whole SNR, where flux densities ($S_{1}$ vs $S_{2}$) are plotted instead of brightness temperatures. Four frequency pairs are included: 44--94 GHz ({\it green hollow squares}); 44--102 GHz ({\it solid black squares}); 102--213 GHz ({\it solid blue squares}); 213--314 GHz ({\it solid red squares}).  Data points for the 44--94 GHz frequency pair  are not fitted.  The given spectral indices are derived from different linear interpolations: least-square fit from vertical squared errors, $\alpha_{y}$ ({\it blue}); orthogonal distance regression, $\alpha_{ODR}$ ({\it green});  robust fitting from the square root of absolute residuals, $\alpha_{Abs}$ ({\it magenta}); linear regression with forced zero interception $(S_{1}=0,S_{2}=0)$, $\alpha_{(0,0)}$ ({\it red}). \\}
\label{fig:TT_Tot}
\end{center}
\end{figure} 

\vspace{-7.0mm}
The spectral index distribution across the remnant is investigated via multi-frequency spectral index maps and the {\it T-T}  plot method (\citealp{cos60,tur62}).
 
Spectral index maps are derived from images at 44, 94, 102, 213 and 345 GHz. In Figure~\ref{fig:spectral_map}, the maps resulting from the combination of images at two, three, four and five frequencies are shown. All images are reduced with identical procedure and restored with a circular beam of $0\farcs8$. ATCA data at 44 and 94 GHz are scaled to day 9280, via exponential fitting parameters derived for ATCA flux densities from day 8000, as measured at 8.6 and 9 GHz (Zanardo et al., in preparation). While the two-frequency maps are derived from direct division of the flux densities, the maps derived from 3, 4 and 5 frequencies are more accurate, as the solution minimizes the function  
\begin{equation}
{\boldsymbol  \varepsilon}_{pq} =\mathlarger{\sum_{k=1}^{n}}\,{ \abs{ \,S_{\nu_{k}}^{^{\,pq}}(\alpha_{M}^{^{pq}}) -  S_{\nu_{0}}^{^{\,pq}} \biggl ( {\nu_{k} \over \nu_{0} } \biggr )^{\alpha_{M}^{^{pq}}} } }^{2}  
\label{eq:Alpha1} 
\end{equation}
at each pixel of coordinates $(p,q)$, for images at $\nu_{k}$ frequencies, with $n\ge3$.

In Figure~\ref{fig:spectral_map}, it can be seen that the spectral indices overall become flatter as the frequencies reach the FIR.
For all maps, the spectral index, $\alpha_{M}$, varies between $-1.25$ and $0.25$, while in the $5-$frequency map the index range narrows to $-0.96\le\alpha_{M}\le0.18$. 
In most of the multi-frequency maps, $\alpha_{M}$ is steeper on the eastern half of the SNR, while flat spectral indices surround the center, where the bulk of dust sits (see \S ~\ref{morph}), and extend onto the NW and SW regions of the remnant. With reference to the $5-$frequency map, $-0.9\lesssim\alpha_{M}\lesssim-0.6$ on the eastern lobe and $-0.4\lesssim\alpha_{M}\lesssim0$ on the western side of the SNR, with $\alpha_{M}\sim0$  around the central region and $-0.4\lesssim\alpha_{M}\lesssim-0.3$ predominantly in the NW quadrant 
(PA $\sim330^{\circ}$) and in the SW quadrant   (PA $\sim210^{\circ}$).		%

The {\it T-T}  plot method is applied to two images at different frequencies, each obtained with identical reduction process and with the same angular resolution. The spectral variations are assessed over image regions not smaller than the beam size, where the spectral index, $\alpha_{TT}$ , is determined from the flux density slope $m$, where $S_{2} = m \,S_{1} + q$.
The regions used for the {\it T-T}  plots are shown in Figure~\ref{fig:TT_boxes_color}. Six frequency pairs are considered: 44$-$102 GHz, 102$-$213 GHz, 213$-$345 GHz; and  44$-$94 GHz, 44$-$213 GHz and 44$-$345 GHz. 
Different box sizes are used to suit the controlling beam of each frequency pair. 
All derived $\alpha_{TT}$  values are listed in Table~\ref{tab_alpha}.

Similarly to the trend of $\alpha_{M}$, $\alpha_{TT}$  values  become flatter at higher frequencies (see Figure~\ref{fig:TT_Tot}).
From 102 to 213 GHz, $\alpha_{TT} \approx-0.1$ across the whole remnant (see Figure~\ref{fig:TT_boxes_color}), while from 213 to 345 GHz the spectral distribution appears markedly split in two larger regions (see also the two-frequency map in Figure~\ref{fig:spectral_map}), with very flat indices on the western half of the SNR and steep indices on the eastern side.  
For the frequency pairs 102--213 and 213--345 GHz, the {\it T-T}  plots for the eastern region (see E in Table~\ref{tab_alpha})
give $\Delta\alpha_{TT} =\alpha_{(102,\,213)}-\alpha_{(213,\,345)}=1.0\pm0.07$. This could be indication of a local spectral break at 213 GHz.
Using $\nu_{c}=213$ GHz in Eq.~\ref{eq:Asym4}, for $B=20$ mG,  $\tau_{e}\sim35$ yr. 
However, given the high Mach number ($\mathcal{M}\sim10^{4}$)
of the eastbound shocks  \citep{zan13}, likely sub-diffusive particle transport \citep{kir96} by the shock front and, consequently, local magnetic-field amplifications \citep{bel04}, it is possible that the CR in the eastern lobe, radiating at $\nu_{c}$, are already past their synchrotron lifetime.

Flat spectral indices in the western lobe extend north and south at both 213--345 GHz and 44--213 GHz (see Figure~\ref{fig:TT_boxes_color}), while {\it T-T}  plots from higher resolution images show a narrower north-south alignment of the flat regions 
(see N1, C1 and W2 in Figure~\ref{fig:TT_boxes_color} and Table~\ref{tab_alpha}),
with $-0.5\lesssim\alpha_{TT} \lesssim-0.3$. 
The {\it T-T}   plots for 44--94 GHz and 44--345 GHz also yield the steepest spectral indices in region W1, this might be due to the local emission drop in the NW sector of the ER, visible in the images at 94 and 345 GHz at PA $\sim300^{\circ}$.

As discussed in \S~\ref{SED},  
the flat-spectrum western regions could be linked to a PWN.
We note that  spectral maps of the Crab Nebula via observations centered at 150 GHz  \citep{are11} have identified  spectral indices around $-0.2$ in the inner central regions of the PWN, while spectral indices $\sim-0.3$ have been associated with the PWN periphery.\\

\section{PWN constraints}
\label{PWN--NS}

As discussed in \S~\ref{Subtract}, the emission at 102 and 213 GHz in the $I_{44}-$ subtracted images appears to peak west of the SN site \citep{rey95} (Figures~\ref{fig:df}, \ref{fig:super}), and to mainly extend west of the optical ejecta (see Figure~\ref{fig:B3679_Hbl}). Besides, both the spectral maps and {\it T-T}  plots (Figures~\ref{fig:spectral_map}, \ref{fig:TT_boxes_color})  show that flat spectral indices can be associated with the center-west regions of the SNR (see \S~\ref{SI}). 
These results could be explained by a possible PWN, powered by a pulsar likely located at a westward offset from the SN position.


The pulsar-kick mechanism has been linked to asymmetries in the core collapse or in the subsequent supernova explosion, presumably due to asymmetric mass ejection and/or asymmetric neutrino emission (\citealp{pod05,won13, nor12}).
As evidence for the  natal kick, neutron star (NS) mean three-dimensional (3D) speeds have been estimated at $\overline{v}_{\rm NS}\sim400\pm40$ km s$^{-1}$ \citep{hob05},
while  a transverse velocity of $\sim1083$ km s$^{-1}$ has been detected  by \citet{cha05}, which would imply a 3D NS birth velocity as high as 1120 km s$^{-1}$ \citep{cha05}.
In the context of SNR 1987A, by day 9280 the NS could have travelled westwards of the SN site by $\sim20-80$ mas, while for an impulsive kick of the same order as  $\overline{v}_{\rm NS}$ a distance of $\sim42\pm5$ mas would have been covered.
With a western offset of $\sim0\farcs05$, the NS would be situated inside the beam of the $I_{44}-$subtracted images associated with the peak flux density, both at 102 and 213 GHz. 
If we take into account the $\sim$60 mas uncertainty intrinsic to image alignment (see \S~\ref{Obs}), as well as the error of  30 mas in each coordinate of the SN VLBI position \citep{rey95}, the NS could be located near the emission peak as seen in the $I_{44}-$subtracted images at 102 and 213 GHz. We note that the inner feature of fainter emission detected in the SNR at 44 GHz, as aligned with VLBI observations of the remnant \citep{zan13}, is centered $\sim60$ mas west of the SN site.


If a pulsar is embedded in the  unshocked ejecta, the PWN would be in its early stages of evolution,  likely surrounded by uniformly expanding gas  \citep{che92}.
Diffuse synchrotron emission from the PWN would be due to the relativistic particles, produced  by the pulsar,  accelerated 
at the wind termination shock \citep{kir09}.

Assuming a power-law energy distribution of electrons, i.e. the particle density $N_{e}$ is expressed as 
$N_{e}(E)\propto K E^{-s}$, where $s = 1-2\alpha$ and $K\propto(m_{e} c^{2})^{s-1}$,
 the synchrotron emission of the PWN can be written as 
\begin{equation}
S_{\nu_{P}} \propto K B_{_{\rm{PWN}}} ^{^{\scalebox{0.9}{$ {1\over2} (s+1) $} }}  \nu\,^{^{\scalebox{0.9}{$ {1\over2} (1-s) $} }}   
\label{eq:PWN1}
\end{equation}
where $B_{_{\rm PWN}}$ is the nebular magnetic field strength. 
Noting that, in radio observations, the energy in electrons cannot be separated from that in the magnetic field \citep{rey12}, 
the equipartition magnetic-field strength could be derived as (e.g. \citealp{lon11};  
see revised formula by \citealp{bec05,arb12})
\begin{equation}
B_{_{\rm{PWN}}} \approx \biggl[ G_{0} \,G \,(\mathcal{K}+1) \, { S_{\nu} \over f d\, \theta_{_{\rm PWN}}^{3}} \,\,
\nu^{\scalebox{1.1}{$ {(1-s)\over2} $} } 
 \biggr]  ^{\scalebox{1.1}{$ { 2 \over (5+s)} $}}
\label{eq:PWN2}
\end{equation}
where $G_{0}$ is a constant, $G=G(\nu,s)$ is the product of different functions varying with the minimum and maximum frequencies associated with the spectral component and the synchrotron spectral index  (\citealp{bec05,lon11}), 
$\mathcal{K}$ is the ion/electron energy ratio, 
$f$ is the volume filling factor of radio emission, and 
$\theta_{_{\rm PWN}}=R_{_{\rm PWN}}/d$ is the angular radius. 
Considering $0\farcs05\lesssim R_{_{\rm PWN}}\lesssim0\farcs15$, 
 $-0.4\le\alpha\le-0.1$ (as from \S~\ref{SED}),
$102\le\nu\le672$ GHz, 
$S_{\nu}\approx 3$ mJy, and taking $\mathcal{K}\approx100$ \citep{bec05} while $f\approx0.5$,  
Eq.~\ref{eq:PWN2} leads to 
 $1\lesssim B_{_{\rm{PWN}}}\lesssim 7$ mG. 
For $0.2\lesssim f\lesssim 0.9$, $2\lesssim B_{_{\rm{PWN}}}\lesssim5$ mG. 
Since the equipartition is a conjecture for young SNRs and no longer valid when the spectral index is flatter than $-0.5$,  these $B_{_{\rm PWN}}$ estimates might be inaccurate.

The energy inside the PWN, 
due to the PWN magnetic field, can be simplified as
\begin{equation}
E_{_{{\rm PWN}, B}}(t)\sim V_{_{\rm PWN}}(t)\frac{B^{2}_{_{\rm PWN}}}{8\pi},
\label{eq: Espin}
\end{equation}
where the magnetic field is considered uniform and isotropic inside the PWN  volume, $V_{_{\rm PWN}}(t)=4/3\, \pi \,R_{_{\rm{PWN}}}^{3}(t)$.
For  
$0\farcs05\lesssim R_{_{\rm{PWN}}}\lesssim 0\farcs15$, 
and,  as for the parameters used in Eq.~\ref{eq:PWN2},
$1\lesssim B_{_{\rm{PWN}}}\lesssim 7$ mG,
at $t=t_{25}=8.0\times 10^{8}$ s ($\approx9280$ days), 
we estimate 
$0.9\times 10^{43}\lesssim E_{_{{\rm PWN}, B}}\lesssim 1.2 \times 10^{46}$ erg.
According to models by  \citet{che92}, about $30\%$ of the total energy input  into the PWN, $E_{_{\rm PWN}}$, goes to the internal magnetic pressure in the PWN, while
most of the remaining pulsar spin-down energy would drive the PWN expansion into the ejecta.

In terms of integrated radio  luminosity, calculated as
\begin{equation}
L_{\rm rad}(\nu)=4 \pi d\,^{2} \int_{\nu_{\rm min}}^{\nu_{\rm max}}\,S_{\nu}(\nu)\, \mathrm{d}\nu,
\label{eq:LPWN}
\end{equation} 
if $\nu_{\rm min}=102$ GHz and $\nu_{\rm max}=672$ GHz bracket the frequency range in which the PWN is detected,
$S_{\nu}\approx3$ mJy leads to $L_{\rm rad}\approx5.4\times10^{33}$ erg s$^{-1}$.
The derived $L_{\rm rad}$  is comparable with the  limit  of  $L_{\rm opt}\le 5 \times10^{33}$ erg s$^{-1}$ given by \citet{gra05} for a compact source in the optical band from 290 to 965 nm at $t= 6110$ days.
A similar limit has been placed on the 2--10 keV X-ray luminosity, $L_{\rm X} \le 5.5 \times 10^{33}$ erg s$^{-1}$, using {\it Chandra} images \citep{par04}.
Since these $L_{\nu}$  estimates are upper limits and free-free 
emission may be a substantial component of the radio luminosity
(see \S~\ref{ff}), we can take $L_{\rm rad}\approx10^{33}$~erg~s$^{-1}$ as a realistic upper limit.

In the free-expansion regime \citep{che92}, $\sim1$\% of the pulsar power is emitted by the shock wave in the supernova and additional emission from the pulsar nebula is expected. 
Given this,
we can set  $\dot{E}_{\rm spin} \approx 10^{35}$~erg~s$^{-1}$ as an upper limit for the spin-down power  of the pulsar.
For a typical pulsar surface dipole magnetic field 
$B_{s} \sim 10^{12}$~G \citep{man05b}, this spin-down
luminosity corresponds to a pulsar period $P\sim0.15$~s and
characteristic age $\tau_{c}= P/(2 \dot{P})  \sim 10^{5}$~years.
Lower luminosities would imply
lower dipole magnetic fields and/or longer pulsar periods. Such parameter
ranges are plausible for the putative pulsar at the centre of SN
1987A, as there is good evidence that many pulsars are born with a
spin period not much different to their present period (\citealp{pop12,got13}).

Following \citet{che77},  the velocity at the outer edge of the nebula can be defined as 
\begin{equation}
v_{_{\rm PWN}}={6 \over 5} \biggl({125 \over 132} \,\,{\dot{E}_{\rm spin}\, t \over \pi \rho_{ej} \, t^{3}} \biggr)^{\scalebox{0.9}{$ { 1 \over 5} $}}, 
\end{equation}
assuming the PWN is freely expanding and has constant density $\rho_{ej}$.
At $t=t_{25}$, setting $10^{-18}\lesssim \rho_{ej}\lesssim 10^{-19}$  g cm$^{-3}$ 
in the central region of the SNR,  as from the density model by  \citet{bli00} (see Figure 21 in \citealp{fra13}), 
the swept-up shell velocity becomes
 $260\lesssim v_{_{\rm PWN}}\lesssim410$ km s$^{-1}$, which 
leads to $R_{_{\rm PWN}}$ not greater than $0\farcs05$. 
Since this is well below the resolution of the ATCA and ALMA images presented here,  the emission from a
possible PWN would appear as a point source.
 

As mentioned in \S~\ref{SED}, a pulsar embedded in the SNR interior would 
emit ionizing radiation within the inner layers of the ejecta. 
While an X-ray pulsar has  yet  to be detected \citep{hel13},
illumination of the inner ejecta by X-ray flux has been reported by \citet{lar11}
though attributed to the reverse and reflected shocks, as well as to
shocks propagating into the ER.

%
%
\begin{figure}[t] 
\begin{center}
\vspace{0.25mm}
\advance\leftskip-5.0mm
\includegraphics[trim=0mm 1.35mm 1.0mm 0.0mm, clip=true,width=89.55mm, angle=0]{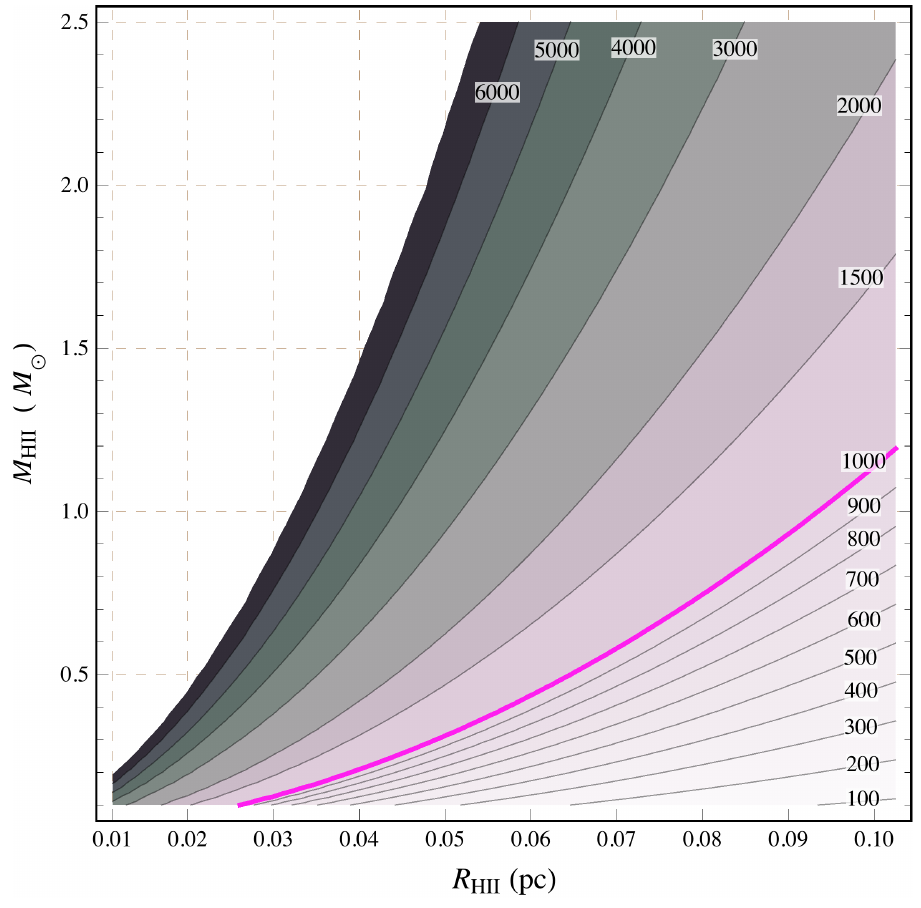}
\caption{
Dispersion measure (DM) associated with a possible pulsar in SNR 1987A, assuming an ionized fraction of the ejecta 
$0.1\,M_{\sun}\lesssim M_{\rm{HII}}\lesssim 2.5\,M_{\sun}$, 
uniformly  distributed within a spherical region with radius
$0.01\lesssim R_{\rm HII} \lesssim 0.10$ pc (i.e.  $0\farcs05\lesssim \ R_{\rm HII} \lesssim 0\farcs42$).
The color scheme changes 
from white to black
for increasing DM values, which range from 100 to 6000 cm$^{-3}$ pc, as indicated by the contour labels. 
The contour at DM $=$1000 cm$^{-3}$ pc is highlighted in magenta. \\ 
}
\label{fig:DM}
\end{center}
\end{figure} 
%
%


Given the early stages of a possible PWN,
the ionized ejecta would be mainly due to the radiation from the various shocks.
If the H{\sc ii} region is assumed to be spherical (see \S~\ref{SED}),
the related dispersion measure (DM) can be derived as
$\,{\rm DM} (M_{\rm HII}, R_{\rm HII})\approx 3 \, M_{\rm HII}/ (m_{p} \, 4\pi \, R_{\rm HII}^{2})$.
Neglecting clumping in the ejecta, 
for $0.1\,M_{\sun}\lesssim M_{\rm{HII}}\lesssim 2.5\,M_{\sun}$ and $0.01\lesssim R_{\rm HII} \lesssim 0.10$ pc (i.e.  $0\farcs05\lesssim \ R_{\rm HII} \lesssim 0\farcs42$),
the resulting DM  is shown in Figure~\ref{fig:DM}.\\

\section{Conclusions}
\label{End}

We have presented a comprehensive morphological and spectral analysis of SNR 1987A based on both ATCA and ALMA data (Cycle 0), from 1.4 to 672 GHz. 
We have investigated the components of the SNR emission across the transition from radio to FIR, as
the combination of non-thermal and thermal emission. A summary of our findings is as follows:

\begin{enumerate}
\item{
To decouple the non-thermal emission from that originating from dust, the synchrotron component, as resolved with ATCA at 44 GHz, and the dust component, as imaged with ALMA at 672 GHz, were subtracted from the datasets at 94, 102, 213, 345 and 672 GHz. The images derived from the subtraction of the scaled model flux density at 672 GHz, highlight the ring-like synchrotron emission that originates over the ER. The images similarly obtained from the subtraction of the 44 GHz model, show residual emission mainly localized west of the SN site.
}
\item{
Analysis  of the emission distribution over the ER, in images from 44 to 345 GHz, 
shows a gradual decrease of the east-to-west asymmetry ratio with frequency. 
We attribute this to the shorter synchrotron lifetime at high frequencies. 
We estimate that, at frequencies higher than 213 GHz, the electrons might be unable to cross 
the eastern emission region.
}
\item{
Across the transition from radio to FIR, the SED 
suggests additional emission components beside the synchrotron 
main component ($S_{\nu}\propto\nu^{-0.73}$) and the thermal emission originating from dust grains at $T\sim 22$ K. 
We argue that this excess emission 
could be due to a second flat-spectrum synchrotron component with $-0.4\lesssim\alpha\lesssim-0.1$.
This could imply the presence of a PWN originating from 
an embedded pulsar.
}
\item{
Spectral index measurements across the SNR, from 44 to 345 GHz, show predominantly flat spectral indices, $-0.4\lesssim\alpha\lesssim-0.1$, in the western half of the remnant at frequencies above 102 GHz, while $\alpha\sim0$ around the central region. From 102 to 213 GHz, $-0.1\lesssim\alpha\lesssim0$ across the whole remnant. 
From 213 to 345 GHz, the steepening of the spectral indices over the eastern lobe might be indication of a local spectral break. 
}
\item{
Results from both the morphological and spectral analysis might suggest  the presence of a PWN in the SNR interior,
powered by a pulsar likely located at a westward offset from the SN position. 
In this scenario, a NS embedded in the unshocked ejecta 
might be emitting diffuse synchrotron emission $S_{\nu_P} \sim 3$ mJy above 102 GHz.  
The possible PWN would have magnetic field strength $1\lesssim B_{_{\rm{PWN}}}\lesssim 7$ mG.
The upper limit of the integrated radio luminosity is derived as
$L_{rad}\sim5.4\times10^{33}$ erg s$^{-1}$, which leads to an upper limit of the pulsar spin-down power $\dot{E}_{\rm spin} \approx 10^{35}$~erg~s$^{-1}$.
While effects such as luminosity, beaming, scattering and absorption might have prevented the detection of pulsed emission  in the SNR for  over 20 years (\citealp{man88, man07}),
a renewed pulsar search with the Parkes telescope is currently in progress. 
Future observations with ALMA, ATCA and VLBI 
are needed 
to probe the SNR for a possible PWN.
}
\end{enumerate}

%
\acknowledgments  
\vspace{-2.0mm}
We thank Philipp Podsiadlowski  and Bruno Leibundgut  for useful discussions;  
Richard McCray, Craig Wheeler and Eli Dwek for feedback on the drafts.
We thank Robert Kirshner and Peter Challis for providing the {\it HST} images.
Figures~\ref{fig:spectral_map} and \ref{fig:TT_boxes_color}
utilize the cube helix color scheme introduced by \citet{gre11}.  
This paper makes use of the following ALMA data: ADS/JAO.ALMA \#2011.0.00273.S (PI: Indebetouw). ALMA is a partnership of ESO (representing its member states), NSF (USA) and NINS (Japan), together with NRC (Canada) and NSC and ASIAA (Taiwan), in cooperation with the Republic of Chile. The Joint ALMA Observatory is operated by ESO, AUI/NRAO and NAOJ. 
The Australia Telescope Compact Array is part of the Australia Telescope, which is funded by the Commonwealth of Australia for operation as a National Facility managed by CSIRO.
L. S-S. and B. M. G.  acknowledge the support of CAASTRO, through project number CE110001020. R.A.C. acknowledges the support of NASA grant NNX12AF90G. 
M. L. acknowledges an ESO/Keele University studentship.

\bibliographystyle{apj}

\clearpage

\end{document}